%% file: towards-cleaned.tex
\RequirePackage[l2tabu, orthodox]{nag}
\RequirePackage{snapshot}

\documentclass[9pt,onecolumn]{extarticle}

\makeatletter
\if@twocolumn
  \usepackage[dvips,letterpaper,top=0.5in, bottom=0.5in, left=0.75in, right=0.5in,includefoot,heightrounded]{geometry}
\else
  \usepackage[dvips,letterpaper,margin=1in,includefoot,heightrounded]{geometry}
\fi

\usepackage{srcltx}

\usepackage[russian,portuges,english]{babel}

\iflanguage{portuges}
    {\newcommand{\keywordname}{Palavras-chaves}}
    {\newcommand{\keywordname}{Keywords}}

\usepackage{amsmath}
\usepackage{amssymb,amsfonts}

\usepackage{abstract}

\usepackage{cite}
\usepackage{amsmath,amssymb,amsfonts}
\usepackage{algorithmic}
\usepackage{graphicx}
\usepackage{textcomp}
\usepackage{cite}
\usepackage{amsmath,amssymb,amsfonts}
\usepackage{algorithmic}
\usepackage{graphicx}
\usepackage{textcomp}

\usepackage{bm}
\usepackage{subfigure}
\usepackage{soul}
\usepackage[english]{babel}

\usepackage{booktabs}
\usepackage{multirow}

\usepackage{psfrag}
\usepackage{epsfig}
\usepackage{flushend}
\usepackage{color}

\usepackage{setspace}

\usepackage{url}
\usepackage{hyperref}

\usepackage[short,12hr]{datetime}

\usepackage{mathtools}

       \usepackage{fouriernc}


\def\BibTeX{{\rm B\kern-.05em{\sc i\kern-.025em b}\kern-.08em
    T\kern-.1667em\lower.7ex\hbox{E}\kern-.125emX}}

\newcommand{\printtitle}{%
\makeatletter
\if@twocolumn

\twocolumn[%
  \maketitle
  \begin{onecolabstract}
    \myabstract
  \end{onecolabstract}
  \begin{center}
    \small
    \textbf{\keywordname}
    \\\medskip
    \mykeywords
  \end{center}
  \bigskip
]
\saythanks
\else
  \maketitle
  \begin{onecolabstract}
    \myabstract
  \end{onecolabstract}
  \begin{center}
    \small
    \textbf{\keywordname}
    \\\medskip
    \mykeywords
  \end{center}
  \bigskip
  \onehalfspacing
\fi
\makeatother
}

\author{%
Arjuna Madanayake%
\thanks{Department of Electrical and Computer Engineering, Florida International University, Miami FL, USA. e-mail: \{amadanay,pberu002,spuli009\}@fiu.edu}
\and
Viduneth Ariyarathna${}^\ast$
\and
Suresh Madishetty%
\thanks{Department of Electrical and Computer Engineering, University of Akron, Akron OH, USA.}
\and
Sravan Pulipati${}^\ast$
\and
R. J. Cintra%
\thanks{Signal Processing Group, UFPE, Brazil. e-mail: {rjdsc@de.ufpe.br}}
\and
Diego Coelho%
\thanks{Independent researcher, Calgary, AB, Canada.
e-mail:~diegofgcoelho@gmail.com}
\and
Ra\'{\i}za Oliveira${}^\ddagger$
\and
F\'abio~M.~Bayer%
\thanks{Departamento de Estat\'{\i}stica and LACESM, UFSM, Brazil. e-mail: bayer@ufsm.br}
\and
Leonid Belostotski%
\thanks{Department of Electrical and Computer Engineering, University of Calgary, Calgary AB, Canada. e-mail:~lbelosto@ucalgary.ca}
\and
Soumyajit Mandal%
\thanks{Department
of Electrical Engineering and Computer Science, Case Western Reserve University, Cleveland OH, USA. e-mail: sxm833@case.edu}
\and
Theodore S. Rappaport%
\thanks{NYU WIRELESS, Department of Electrical and Computer Engineering at New York University Tandon School of Engineering, New York NY, USA. e-mail: tsr@nyu.edu}
}

\title{%
Towards a Low-SWaP 1024-beam Digital Array: A 32-beam Sub-System at 5{.}8~GHz}

\newcommand{\myabstract}{%
Millimeter wave communications require multibeam beamforming in order to utilize wireless channels that suffer from obstructions, path loss, and multi-path effects. Digital multibeam beamforming has maximum degrees of freedom compared to analog phased arrays. However, circuit complexity and power consumption are important constraints for digital multibeam systems. A low-complexity digital computing architecture is proposed for a multiplication-free 32-point linear transform that approximates multiple simultaneous RF beams similar to a discrete Fourier transform (DFT).
Arithmetic complexity due to multiplication is reduced from the FFT complexity of $\mathcal{O}(N\: \log N)$ for DFT realizations, down to zero, thus yielding a 46\% and 55\% reduction in chip area and dynamic power consumption, respectively, for the $N=32$ case considered.
The paper describes the proposed 32-point DFT approximation targeting a 1024-beams using a 2D array, and shows the multiplierless approximation and its mapping to a 32-beam sub-system consisting of 5.8 GHz antennas that can be used for generating 1024 digital beams without multiplications.
Real-time beam computation is achieved using a Xilinx FPGA at 120 MHz bandwidth per beam. Theoretical beam performance is compared with measured RF patterns from both a fixed-point FFT as well as the proposed multiplier-free algorithm and are in good agreement.
}

\newcommand{\mykeywords}{%
Approximate beamforming, multibeams, digital arrays.
}

\date{}

\begin{document}

\printtitle

\section{Introduction}
\label{sec:introduction}

The efficient formation of far-field antenna patterns simultaneously across a multitude of directions is crucially important for wireless communications, radio astronomy,  imaging, radar, and electronic warfare. Multibeam beamforming has been usually achieved in the microwave domain using analog techniques (e.g., Rotman lenses~\cite{cmbref6} and Butler/Nolan matrices~\cite{cmbref26,cmbref31}). Emerging mmW systems are considering hybrid multibeam beamforming due to its power efficiency and excellent performance for a reasonably small number of antenna elements and user streams~\cite{mmWHBF_SunRapShafi,hybridSun2}.
Although digital beamforming requires
the
control of each individual antenna element in an antenna array,
it is promising for the future due to its many inherent advantages, which include~\cite{yeary2016}: i) maximum flexibility/reconfigurabilty; ii) easy system updates and support for new beamforming algorithms as they emerge; iii) precise control of both the gain and phase of individual antenna elements thus giving better control of the beams; iv) maximum degrees of freedom from a given array; and v) reduced maintenance and calibration requirements.

Element-wise digital beamforming requires a dedicated receiver (or transmitter, in transmit mode) for each antenna element,
which
is
usually a uniformly spaced linear or rectangular array of antennas.
Multibeams can be generated by expanding the concept of a phased-array to multiple simultaneous directions by using the fact that each direction of propagation of a carrier wave is associated with two spatial frequencies $\left(\omega_x,\omega_y\right)\in\mathbb{R}^2$ across the two orthogonal coordinate axes of a rectangular array aperture. Multiple beam digital beamforming is desired at the lowest possible energy consumption for a given bandwidth, supply voltage, and technology node, which leads to domain-specific architectures that are optimized for low complexity and power consumption.

Here we propose approximate computing-based algorithms and computing architectures that achieve quasi-orthogonal RF beams without using any digital multiplier circuits.
The multiplierless nature of the digital computing architectures allow low chip area/size, weight, and power consumption (SWaP) and avoid the need for digital multipliers that have high circuit complexity (transistor count) and power consumption.
This is likely to become more critical as wireless systems move to sub-terahertz frequencies and much wider channel bandwidths than used currently \cite{6G_ACCESS}. Algorithms that are multiplierless thus lead to substantially reduced SWaP in real-time digital silicon implementations~\cite{LowSWAP_fundermental,6G_ACCESS}.

Multibeam beamforming on linear/rectangular apertures is important for exploiting multi-directional channels in massive-MIMO systems, for example in 5G mm-wave wireless networks.
Such systems rely on the combination of beamforming with MIMO theory~\cite{multibeam_overview,5G_overview_rappaport,Debani_5G_overview,mmWHBF_SunRapShafi,6G_ACCESS,hybridSun2} and as frequencies move to THz ranges, the need for providing thousands of simultaneous beams will emerge due to the small size of the wavelength and physical antenna aperture. Recent work has described different phased array architectures targeting 5G applications~\cite{shu2018,Rabeiz_32_element_17,Flat_Rotman_16,Rebeiz_17,5G_vehicular,Rebeiz_multibeam_digital_2016,Lota2017}. However, most of the literature has been concentrated either on hybrid beamforming systems or fully-analog architectures due to the prohibitive processing complexity of fully-digital beamforming.
Other important applications for microwave and mm-wave digital multibeam beamformers include emerging defense applications such as space-based low earth orbit communications, mesh networks between micro satellites, space-based Internet distribution to densely populated areas, and multi-domain mosaic warfare where reliable high-speed wireless connectivity is needed across multiple platforms (air, space, land, and sea), as well as for emerging terahertz imaging systems \cite{6G_ACCESS}. The demands of high-capacity wireless networks for such applications can be significantly more difficult to meet than commercial 5G standards since modern military platforms can travel at hypersonic speeds across hundreds of kilometers. Such demanding wireless channel conditions necessitate beamforming gain across wide bandwidths and narrow angles of propagation (i.e., sharp beams) to both thwart detection and also benefit from beamforming gain. Furthermore, 5G networks will eventually require
digital beamforming to reduce the overhead associated with the current 3GPP beam search time in the 5G game structure---great reduction in beam pointing can be obtained by simultaneously searching the environment for the best pointing angle, but this is not yet supported in the 5G 3GPP standard~\cite{5G3gpp}.

Some recent work has focused on achieving element-wise fully-digital beamforming. The paper in \cite{Jeong_DBF_2016} presents a low-power 8-element digital beamforming prototype based on bit-stream processing. The design uses a low-resolution $\Delta \Sigma$ architecture that replaces multipliers with multiplexers. This multiplexer-based architecture achieves lower power and smaller area than conventional digital beamformers, but the design is limited to a 20~MHz bandwidth with only two simultaneous output beams.
Another recent paper~\cite{Jang_JSSC_18} reports a 16-element 4-beam digital beamformer targeting large scale MIMO for 5G communications systems. It uses a similar multiplexer-based approach as in \cite{Jeong_DBF_2016} with an interleaved architecture to support a 100~MHz bandwidth.
The work in \cite{5G_Experimental_1,multi-beam-mimo} also report experimental verification of fully digital multibeam beamforming schemes targeting MIMO-based 5G implementations.

The paper~\cite{multibeam-fft-1997} presents a spatial DFT-based digital multibeam beamforming implementation scheme for satellite communications. The earlier work of authors in \cite{ AriyarathnaTAP18} describes a low-complexity algorithm using the spatial DFT based approach for generating 16 simultaneous beams using a ULA. The work in \cite{ AriyarathnaTAP18} uses a 16-point DFT approximation to generate the simultaneous beams and presents the measured beams of its fully digital implementation targetting 5G MIMO applications. This paper describes a novel low-complexity algorithm for realizing a massive number of simultaneous sharp beams, which are vitally important in coping with the rapid increases in path loss expected in future mmW/sub-THz wireless systems. In particular, we propose a low-SWaP approach to generating 1024 beams using a $32\times 32$ aperture and ultra-low-complexity digital VLSI hardware. We propose a 32-beam subsystem based on a novel 32-point DFT approximation as the building block of such a $32 \times 32$ system. The proposed 32-beam subsystem has been implemented at 5.8~GHz and the digital beams have measured and compared with those from exact-DFT-based beams. The measured beams have been used to derive the beam patterns of the corresponding $32\times 32$ rectangular aperture by assuming identical element patterns in all directions.

The paper is structured in as follows. Section~I provides a introduction to the paper %
Section~II, followed by the introduction, describes the role of the DFT for spatial filtering in multibeam transceivers. Section~III describes a 32-point approximate DFT algorithm with zero multiplicative complexity. Section~IV describes an experimental realization of a 32-element receive array that can implement the proposed approximate DFT algorithm for digital beamforming. Experimental results from this array are presented in Section~V. Finally, Section~VI concludes the paper.

\section{Role of the DFT in Multibeam Transceivers}

Multibeam beamforming on an $N\times M$ ($N,~M \in \mathbb{Z^+}$) linearly spaced rectangular array can be achieved by uniformly sampling the spatial frequency domains to define a set of far-field plane-waves having spatial frequencies determined by setting $\left(\omega_x,\omega_y\right)=\left(\frac{2\pi}{N}k_1,\frac{2\pi}{M}k_2\right)$ where $k_1=0,1,\ldots,M-1$ and
$k_2=0,1,\ldots,N-1$.
For this analysis we consider $M =N$ so that the same proposed $N$-point transform can be used row-wise and column-wise in the rectangular aperture for generating two-dimensional (2D) beams. The spatial frequency points $\left(\frac{2\pi}{N}k_1,\frac{2\pi}{N}k_2\right)$ correspond to beams pointing at unique angle pairs indexed by $\left(k_1,k_2\right)\in\mathbb{Z}^2$.
The corresponding spatially-sampled time-continuous plane waves at the terminals of the array elements can be expressed in a Fourier basis as $E(n_x,n_y,t)=E_0 \sum_{k_1=0}^{N-1} \sum_{k_2=0}^{N-1} x_{m}(t)e^{j(n_x\omega_x+n_y\omega_y+2\pi f_ct)}$ where $f_c$ is the unmodulated carrier frequency, $E_0$ is a constant that sets the signal power, and $x_{m}(t)$ is the complex modulated information component of the signal. Note that we assume that the bandwidth of $x_{m}(t)$ is much smaller than $f_c$, since our analysis is only valid for narrowband signals for which the so-called \emph{spatial-wideband effect} is negligible~\cite{liu2010wideband,spatialWidebandEffect_Rap}.

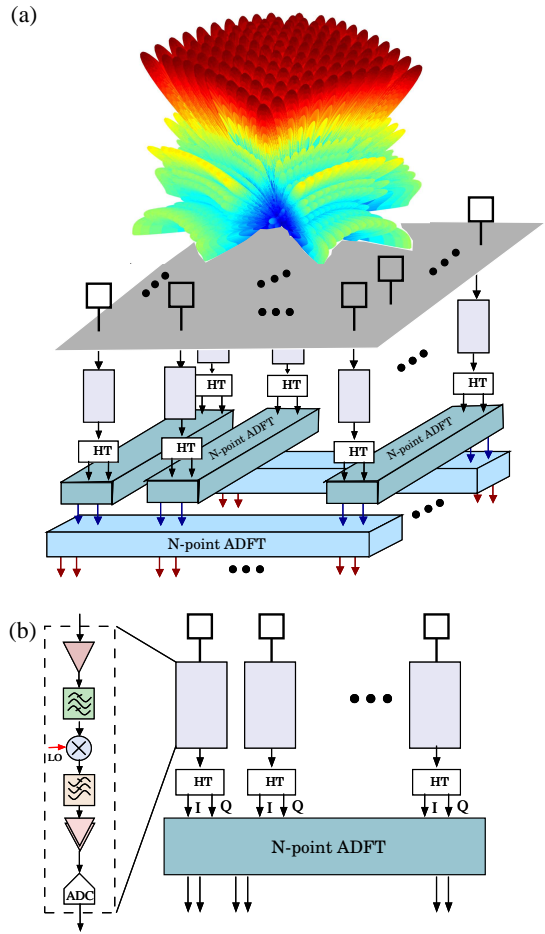
\begin{figure}
	\centering
	\scalebox{1}{\input{32_ULA_URA_OV_1024_2.pstex_t}}
	\caption{(a) Digital beamforming architecture for obtaining $N^2$ beams using an $N \times N$ URA. (b) Block diagram of a $N$-element sub-system  that acts as a building block for the $N^2$ rectangular aperture array. The \textbf{HT} block in the figure denotes the Hilbert transform operation. \label{fig_ULA_URA_archs}}
\vskip -1ex
\end{figure}

In receiver mode, the plane-waves present at the antennas are sampled in the spatial domain, amplified and filtered, down-converted to baseband (or an intermediate frequency (IF), and finally digitized by an analog-to-digital converter (ADC) present at each array location. The digitized signals at each location is complex, i.e., has in-phase ($I$) and quadrature ($Q$) components. For example, down-conversion using a quadrature mixer (which is modeled as multiplication by $e^{-j2\pi f_ct}$) leaves the  spatial frequency components $E_{BB}(n_x,n_y,t)=x_{m,k}(t)e^{j(n_x\omega_x+n_y\omega_y)}$ intact. As a result, the spatial spectrum of the wave remains localized at $\left(\omega_x,\omega_y \right)$. The creation of a sharp RF beam for extracting directional information for a particular plane-wave therefore involves the application of a 2D spatial bandpass filter having the sharpest possible selectivity centered on a particular frequency pair in the spatial frequency domain. From filter theory, it is known that the discrete Fourier transform (DFT) realizes a filterbank of finite impulse response (FIR) filters with sharp bandpass responses that take the well-known
$\operatorname{sinc}(\omega)$ response shape; the peak stopband magnitude for this shape has an asymptote of to $-13.25$~dB for increasing filter order $N$. Therefore, to simultaneously receive an $N\times N$ array of signals, the multibeam beamformer must compute the 2D DFT spatially across the $n_x$-$n_y$ dimensions of the array.

For transmit applications, the waves to be transmitted at simultaneous multiple directions are applied to the inputs of the 2D inverse DFT (IDFT), with the corresponding IDFT outputs being converted to analog using digital-to-analog converters (DACs), filtered, up-converted to the desired carrier frequency, and amplified before being applied to the input terminals of the transmit array. Thus, the computation of the 2D spatial DFT/IDFT for each new sample of the digital baseband signal is a straightforward technique for achieving a large number of simultaneous RF beams in both receive and transmit modes.

Fig.~\ref{fig_ULA_URA_archs}(a) shows the digital beamforming architecture for an $N\times N$ uniform rectangular aperture (URA) that generates $N^2$ beams.
The block diagram of an $N$-element uniform linear array (ULA) subsystem that acts as a building block for the $N^2$ URA is shown in Fig.~\ref{fig_ULA_URA_archs}(b).
The direct computation of the DFT of an~$N$-point
vector of input values is defined as the
matrix-vector multiplication $\mathbf{X}=\mathbf{F}_N\cdot\mathbf{x}$
where $\mathbf{x}$ and~$\mathbf{X}$ are both $N$-point complex column
vectors
and $\mathbf{F}_N$ is the
$N\times N$ complex matrix,
known as the DFT matrix~\cite{Blahut2010}.
The direct matrix-vector multiplication
requires a number of complex arithmetic operation
in $\mathcal{O}(N^2)$,
where~$\mathcal{O}(\cdot)$ is O-notation~\cite[p.~429]{graham1994concrete}.
However, because of the symmetries of the DFT matrix,
it is possible to compute the matrix-vector
product~$\mathbf{X}=\mathbf{F}_N\cdot\mathbf{x}$
with less than $N^2$
complex arithmetic operations.
Algorithms that exploit the
DFT matrix structure are called fast Fourier transforms (FFTs).
FFTs are a famous and well-explored class of
fast algorithms based on sparse matrix factorizations,
which ultimately reduce the
arithmetic complexity to compute~$\mathbf{X}=\mathbf{F}_N\cdot\mathbf{x}$
to be of $\mathcal{O}(N\:\log N)$.
The complexity reduction
from $\mathcal{O}(N^2)$ to $\mathcal{O}(N\:\log N)$
is substantial as $N$ grows large,
which explains the importance of the use of FFTs
in place of the DFT.

If we consider a single-beam conventional digital beamforming scenario, that would need  3$N$ real multiplications and 5$N$ + 2($N$-1) = 7$N$-2 addition operations per beam.
Considering this complexity, if $N$ arbitrary beams are needed, then such system would demand for 3$N^2$ real multiplications and $7N^2- 2N$ real additions. The use of FFTs brings down this complexity to the order of $N\log N$. Using the proposed approach mentioned in section \ref{sec:32pointADFT}, we can obtain FFT-like 32 one-dimensional (1D) beams at no multiplications and with only addition operations. %
We believe that these kind of large number of beams are required for applications like millimeter wave 5G communications as described in the introduction of the paper. %
This method can be an overkill for applications that require only few number of beams. But for applications that do need multiple simultaneous beams, the proposed method provides an attractive solution with a much lower power consumption and area in VLSI implementations.

\section{A 32-point DFT Approximation and Fast Algorithm for RF Beamforming} \label{sec:32pointADFT}

\subsection{Approximate Computing}
The implementation of FFT/DFT in fixed-point digital hardware always leads to errors in representing the twiddle factors~\cite{FFT_twiddle_factor} which are mostly irrational.
Therefore, the implementation of an FFT in physical systems is not perfect and is always an approximation.
If an approximation can be used with better overall performance than that which results from the error sources of the system, then it makes sense to adopt such an approximation with the guarantee that it does not cause other impairments.

The difficulty in proposing DFT approximations for larger sequences
rely on the hardness of the deriving efficient
fast algorithms for generated approximations~\cite{Cintra2011a},
simply because the approximate transforms may not
preserve the same symmetries and mathematical properties
that exist in the exact DFT matrix.

\subsection{32-point Approximate DFT (ADFT)}

Here we describe a 32-point approximate DFT matrix $\mathbf{\hat{F}}_{32}$ for which the matrix-vector multiplication operation can be computed without multipliers.
Let~$\mathbb{P}$ be the set~$\{0, \pm 1, \pm 2, \pm 1/2\}$.
Let~$\mathcal{M}_{\mathbb{P}^2}(32)$
be the set of $32\times 32$
complex matrices
such that the real and the imaginary parts are
defined
over the set~$\mathbb{P}$.
The approximate transform~$\mathbf{\hat{F}}_{32}$
can be found according to a multi-criterion optimization considering
the search space represented
by the parametrized mapping below:
\begin{align*}
g: \mathbb{R} & \longrightarrow \mathcal{M}_{\mathbb{P}^2}(32)
,
\\
\beta & \longmapsto \operatorname{round}(\beta\cdot\mathbf{F}_{32})
\end{align*}
and objective functions given by the following selected
matrix-based metrics:
(i)~Frobenius norm of the matrix difference,
(ii)~total error energy,
(iii)~average percent absolute error,
and
(iv)~orthogonality deviation.
The optimal solution for the DFT approximation can be found
by the
determination of the Pareto efficient solution set,
which is the set of non-dominant solutions~\cite{Ehrgott2005}
using~$\beta \in (0, 5]$ with steps of~$10^{-2}$.

The optimal matrix resulting from the above optimization problem
is given by
\begin{align} \label{eq:adft32}
\mathbf{\hat{F}}_{32} =
\left[
\begin{matrix}%
\mathbf{A}_{0} & \mathbf{A}_{1}\\
\mathbf{A}_{2} & \mathbf{A}_{3}\\
\end{matrix}
\right],
\end{align}
where $\mathbf{A}_i$, $i \in \{0,1,2,3\}$,
are $16\times 16$ sub-matrices given by
equations~\eqref{eq-32pointDFTApproximation-0}--\eqref{eq-32pointDFTApproximation-3}~\cite{DoraThesis}.

\begin{align}
\label{eq-32pointDFTApproximation-0}
\mathbf{A}_{0} =
\left[
\begin{smallmatrix}%
	1 & 1 & 1 & 1 & 1 & 1 & 1 & 1 & 1 & 1 & 1 & 1 & 1 & 1 & 1 & 1\\
	1 & 1 & 1 & 1-1i & 1-1i & 1-1i & -1i & -1i & -1i & -1i & -1i & -1-1i & -1-1i & -1-1i & -1 & -1\\
	1 & 1 & 1-1i & -1i & -1i & -1i & -1-1i & -1 & -1 & -1 & -1+1i & 1i & 1i & 1i & 1+1i & 1\\
	1 & 1-1i & -1i & -1i & -1-1i & -1 & -1 & -1+1i & 1i & 1+1i & 1 & 1 & 1-1i & -1i & -1i & -1-1i\\
	1 & 1-1i & -1i & -1-1i & -1 & -1+1i & 1i & 1+1i & 1 & 1-1i & -1i & -1-1i & -1 & -1+1i & 1i & 1+1i\\
	1 & 1-1i & -1i & -1 & -1+1i & 1i & 1 & 1-1i & -1i & -1-1i & -1 & 1i & 1+1i & 1 & -1i & -1-1i\\
	1 & -1i & -1-1i & -1 & 1i & 1 & 1-1i & -1i & -1 & 1i & 1+1i & 1 & -1i & -1 & -1+1i & 1i\\
	1 & -1i & -1 & -1+1i & 1+1i & 1-1i & -1i & -1 & 1i & 1 & -1i & -1-1i & -1+1i & 1+1i & 1 & -1i\\
	1 & -1i & -1 & 1i & 1 & -1i & -1 & 1i & 1 & -1i & -1 & 1i & 1 & -1i & -1 & 1i\\
	1 & -1i & -1 & 1+1i & 1-1i & -1-1i & 1i & 1 & -1i & -1 & 1i & 1-1i & -1-1i & -1+1i & 1 & -1i\\
	1 & -1i & -1+1i & 1 & -1i & -1 & 1+1i & -1i & -1 & 1i & 1-1i & -1 & 1i & 1 & -1-1i & 1i\\
	1 & -1-1i & 1i & 1 & -1-1i & 1i & 1 & -1-1i & 1i & 1-1i & -1 & 1i & 1-1i & -1 & 1i & 1-1i\\
	1 & -1-1i & 1i & 1-1i & -1 & 1+1i & -1i & -1+1i & 1 & -1-1i & 1i & 1-1i & -1 & 1+1i & -1i & -1+1i\\
	1 & -1-1i & 1i & -1i & -1+1i & 1 & -1 & 1+1i & -1i & -1+1i & 1 & -1 & 1+1i & -1i & 1i & 1-1i\\
	1 & -1 & 1+1i & -1i & 1i & -1i & -1+1i & 1 & -1 & 1 & -1-1i & 1i & -1i & 1i & 1-1i & -1\\
	1 & -1 & 1 & -1-1i & 1+1i & -1-1i & 1i & -1i & 1i & -1i & 1i & 1-1i & -1+1i & 1-1i & -1 & 1\\
	\end{smallmatrix}
\right]
,
\end{align}

\begin{align}
\label{eq-32pointDFTApproximation-1}
\mathbf{A}_{1} =
\left[
\begin{smallmatrix}%
	1 & 1 & 1 & 1 & 1 & 1 & 1 & 1 & 1 & 1 & 1 & 1 & 1 & 1 & 1 & 1\\
	-1 & -1 & -1 & -1+1i & -1+1i & -1+1i & 1i & 1i & 1i & 1i & 1i & 1+1i & 1+1i & 1+1i & 1 & 1\\
	1 & 1 & 1-1i & -1i & -1i & -1i & -1-1i & -1 & -1 & -1 & -1+1i & 1i & 1i & 1i & 1+1i & 1\\
	-1 & -1+1i & 1i & 1i & 1+1i & 1 & 1 & 1-1i & -1i & -1-1i & -1 & -1 & -1+1i & 1i & 1i & 1+1i\\
	1 & 1-1i & -1i & -1-1i & -1 & -1+1i & 1i & 1+1i & 1 & 1-1i & -1i & -1-1i & -1 & -1+1i & 1i & 1+1i\\
	-1 & -1+1i & 1i & 1 & 1-1i & -1i & -1 & -1+1i & 1i & 1+1i & 1 & -1i & -1-1i & -1 & 1i & 1+1i\\
	1 & -1i & -1-1i & -1 & 1i & 1 & 1-1i & -1i & -1 & 1i & 1+1i & 1 & -1i & -1 & -1+1i & 1i\\
	-1 & 1i & 1 & 1-1i & -1-1i & -1+1i & 1i & 1 & -1i & -1 & 1i & 1+1i & 1-1i & -1-1i & -1 & 1i\\
	1 & -1i & -1 & 1i & 1 & -1i & -1 & 1i & 1 & -1i & -1 & 1i & 1 & -1i & -1 & 1i\\
	-1 & 1i & 1 & -1-1i & -1+1i & 1+1i & -1i & -1 & 1i & 1 & -1i & -1+1i & 1+1i & 1-1i & -1 & 1i\\
	1 & -1i & -1+1i & 1 & -1i & -1 & 1+1i & -1i & -1 & 1i & 1-1i & -1 & 1i & 1 & -1-1i & 1i\\
	-1 & 1+1i & -1i & -1 & 1+1i & -1i & -1 & 1+1i & -1i & -1+1i & 1 & -1i & -1+1i & 1 & -1i & -1+1i\\
	1 & -1-1i & 1i & 1-1i & -1 & 1+1i & -1i & -1+1i & 1 & -1-1i & 1i & 1-1i & -1 & 1+1i & -1i & -1+1i\\
	-1 & 1+1i & -1i & 1i & 1-1i & -1 & 1 & -1-1i & 1i & 1-1i & -1 & 1 & -1-1i & 1i & -1i & -1+1i\\
	1 & -1 & 1+1i & -1i & 1i & -1i & -1+1i & 1 & -1 & 1 & -1-1i & 1i & -1i & 1i & 1-1i & -1\\
	-1 & 1 & -1 & 1+1i & -1-1i & 1+1i & -1i & 1i & -1i & 1i & -1i & -1+1i & 1-1i & -1+1i & 1 & -1\\
	\end{smallmatrix}
\right]
,
\end{align}

\begin{align}
\label{eq-32pointDFTApproximation-2}
\mathbf{A}_{2} =
\left[
\begin{smallmatrix}%
	1 & -1 & 1 & -1 & 1 & -1 & 1 & -1 & 1 & -1 & 1 & -1 & 1 & -1 & 1 & -1\\
	1 & -1 & 1 & -1+1i & 1-1i & -1+1i & -1i & 1i & -1i & 1i & -1i & 1+1i & -1-1i & 1+1i & -1 & 1\\
	1 & -1 & 1-1i & 1i & -1i & 1i & -1-1i & 1 & -1 & 1 & -1+1i & -1i & 1i & -1i & 1+1i & -1\\
	1 & -1+1i & -1i & 1i & -1-1i & 1 & -1 & 1-1i & 1i & -1-1i & 1 & -1 & 1-1i & 1i & -1i & 1+1i\\
	1 & -1+1i & -1i & 1+1i & -1 & 1-1i & 1i & -1-1i & 1 & -1+1i & -1i & 1+1i & -1 & 1-1i & 1i & -1-1i\\
	1 & -1+1i & -1i & 1 & -1+1i & -1i & 1 & -1+1i & -1i & 1+1i & -1 & -1i & 1+1i & -1 & -1i & 1+1i\\
	1 & 1i & -1-1i & 1 & 1i & -1 & 1-1i & 1i & -1 & -1i & 1+1i & -1 & -1i & 1 & -1+1i & -1i\\
	1 & 1i & -1 & 1-1i & 1+1i & -1+1i & -1i & 1 & 1i & -1 & -1i & 1+1i & -1+1i & -1-1i & 1 & 1i\\
	1 & 1i & -1 & -1i & 1 & 1i & -1 & -1i & 1 & 1i & -1 & -1i & 1 & 1i & -1 & -1i\\
	1 & 1i & -1 & -1-1i & 1-1i & 1+1i & 1i & -1 & -1i & 1 & 1i & -1+1i & -1-1i & 1-1i & 1 & 1i\\
	1 & 1i & -1+1i & -1 & -1i & 1 & 1+1i & 1i & -1 & -1i & 1-1i & 1 & 1i & -1 & -1-1i & -1i\\
	1 & 1+1i & 1i & -1 & -1-1i & -1i & 1 & 1+1i & 1i & -1+1i & -1 & -1i & 1-1i & 1 & 1i & -1+1i\\
	1 & 1+1i & 1i & -1+1i & -1 & -1-1i & -1i & 1-1i & 1 & 1+1i & 1i & -1+1i & -1 & -1-1i & -1i & 1-1i\\
	1 & 1+1i & 1i & 1i & -1+1i & -1 & -1 & -1-1i & -1i & 1-1i & 1 & 1 & 1+1i & 1i & 1i & -1+1i\\
	1 & 1 & 1+1i & 1i & 1i & 1i & -1+1i & -1 & -1 & -1 & -1-1i & -1i & -1i & -1i & 1-1i & 1\\
	1 & 1 & 1 & 1+1i & 1+1i & 1+1i & 1i & 1i & 1i & 1i & 1i & -1+1i & -1+1i & -1+1i & -1 & -1\\
	\end{smallmatrix}
\right]
,
\end{align}

\begin{align}
\label{eq-32pointDFTApproximation-3}
\mathbf{A}_{3} =
\left[
\begin{smallmatrix}%
	1 & -1 & 1 & -1 & 1 & -1 & 1 & -1 & 1 & -1 & 1 & -1 & 1 & -1 & 1 & -1\\
	-1 & 1 & -1 & 1-1i & -1+1i & 1-1i & 1i & -1i & 1i & -1i & 1i & -1-1i & 1+1i & -1-1i & 1 & -1\\
	1 & -1 & 1-1i & 1i & -1i & 1i & -1-1i & 1 & -1 & 1 & -1+1i & -1i & 1i & -1i & 1+1i & -1\\
	-1 & 1-1i & 1i & -1i & 1+1i & -1 & 1 & -1+1i & -1i & 1+1i & -1 & 1 & -1+1i & -1i & 1i & -1-1i\\
	1 & -1+1i & -1i & 1+1i & -1 & 1-1i & 1i & -1-1i & 1 & -1+1i & -1i & 1+1i & -1 & 1-1i & 1i & -1-1i\\
	-1 & 1-1i & 1i & -1 & 1-1i & 1i & -1 & 1-1i & 1i & -1-1i & 1 & 1i & -1-1i & 1 & 1i & -1-1i\\
	1 & 1i & -1-1i & 1 & 1i & -1 & 1-1i & 1i & -1 & -1i & 1+1i & -1 & -1i & 1 & -1+1i & -1i\\
	-1 & -1i & 1 & -1+1i & -1-1i & 1-1i & 1i & -1 & -1i & 1 & 1i & -1-1i & 1-1i & 1+1i & -1 & -1i\\
	1 & 1i & -1 & -1i & 1 & 1i & -1 & -1i & 1 & 1i & -1 & -1i & 1 & 1i & -1 & -1i\\
	-1 & -1i & 1 & 1+1i & -1+1i & -1-1i & -1i & 1 & 1i & -1 & -1i & 1-1i & 1+1i & -1+1i & -1 & -1i\\
	1 & 1i & -1+1i & -1 & -1i & 1 & 1+1i & 1i & -1 & -1i & 1-1i & 1 & 1i & -1 & -1-1i & -1i\\
	-1 & -1-1i & -1i & 1 & 1+1i & 1i & -1 & -1-1i & -1i & 1-1i & 1 & 1i & -1+1i & -1 & -1i & 1-1i\\
	1 & 1+1i & 1i & -1+1i & -1 & -1-1i & -1i & 1-1i & 1 & 1+1i & 1i & -1+1i & -1 & -1-1i & -1i & 1-1i\\
	-1 & -1-1i & -1i & -1i & 1-1i & 1 & 1 & 1+1i & 1i & -1+1i & -1 & -1 & -1-1i & -1i & -1i & 1-1i\\
	1 & 1 & 1+1i & 1i & 1i & 1i & -1+1i & -1 & -1 & -1 & -1-1i & -1i & -1i & -1i & 1-1i & 1\\
	-1 & -1 & -1 & -1-1i & -1-1i & -1-1i & -1i & -1i & -1i & -1i & -1i & 1-1i & 1-1i & 1-1i & 1 & 1\\
	\end{smallmatrix}
\right]
.
\end{align}

Among the efficient solutions,
the matrix~$\mathbf{\hat{F}}_{32}$
exhibits
the smallest total error energy of approximately~$3.32\cdot 10^{2}$.
The Frobenius norm of the matrix error \emph{per matrix element}
$
\frac{1}{32^2}
\|
\mathbf{\hat{F}}_{32}-\mathbf{{F}}_{32}
\|_F
=
1.004\cdot 10^{-2}$,
where $\| \cdot \|_F$
is the Frobenius norm.
This measurement is
54.9\% lower
than the error per element
of the DFT approximation described in~\cite{Suarez2014,Coutinho2018a,AriyarathnaTAP18}
and can be regarded as acceptable.

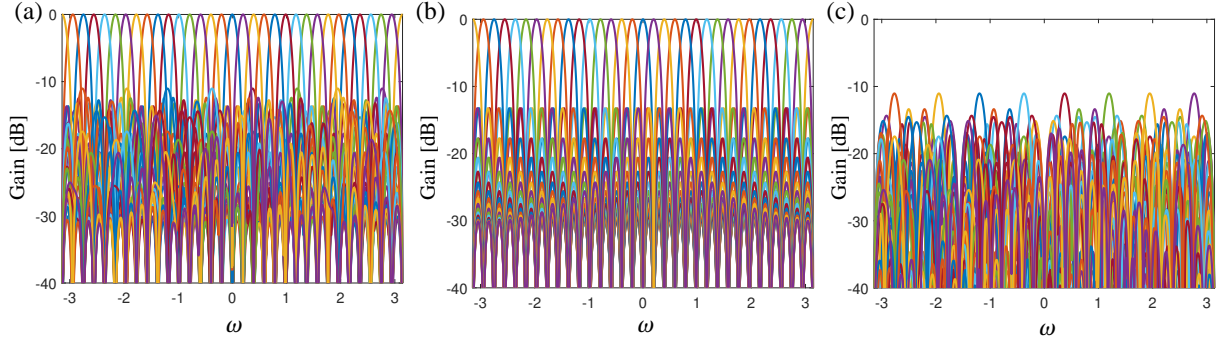
\begin{figure*}
\centering
\scalebox{0.8}{\input{Fig_sim_algo1.pstex_t}}

\caption{The simulated frequency responses of the 32 output bins of the
(a)~proposed 32-point ADFT,
(b)~exact DFT;
(c)~the magnitude error of the two responses.}
\label{fig_sim_algo1}

\end{figure*}

\begin{figure*}
\centering
\scalebox{0.9}{
\input{worst_bins.pstex_t}
}
	\caption{Bins that have the highest magnitude error in Fig. \ref{fig_sim_algo1}. (c). \label{fig_worst_bins}}
\end{figure*}
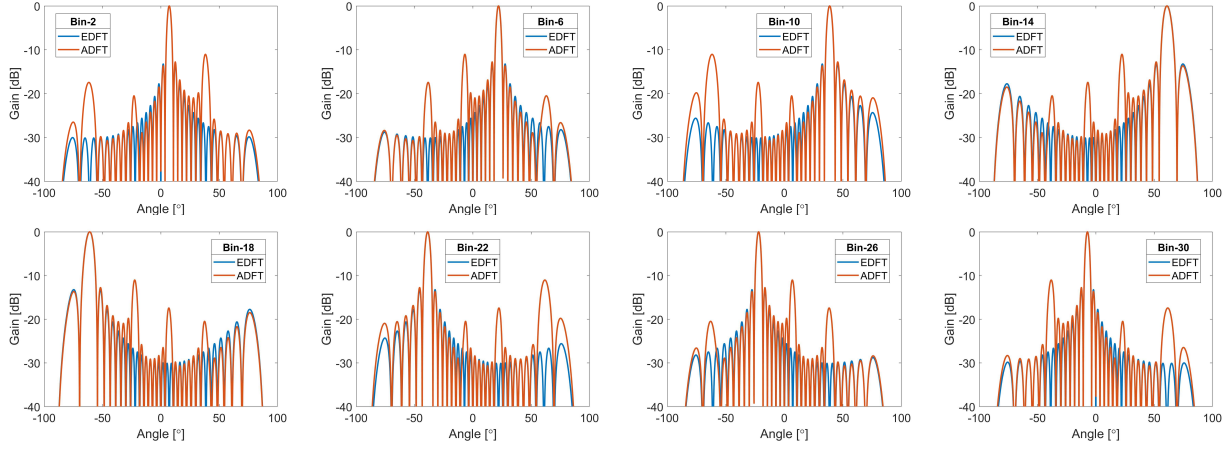

Fig.~\ref{fig_sim_algo1} shows a comparison of the frequency responses of all the bins for the 32-point proposed approximate DFT (ADFT) and the DFT.
The shapes and locations of the main beams are almost identical to the exact DFT. The relative errors of the magnitude response of each filter response are largely confined to the stopbands away from the main lobe (i.e., deep side lobes), and are generally below the $-15$~dB level. Fig.~\ref{fig_sim_algo1}(c) shows the magnitude error plot of the filter bank responses of the proposed DFT approximation. The plot in Fig.~\ref{fig_sim_algo1}(c) is computed by evaluating the difference of the magnitude responses of approximate and exact DFT transforms for each filter (i.e., DFT/ADFT bin). The plots in Fig.~\ref{fig_worst_bins} show the bins in Fig.~\ref{fig_sim_algo1}(c) that have the highest magnitude error. All other bins have a magnitude error that is smaller than~$-13$ dB. The deviations in the filter bank responses with respect to
the DFT filter bank responses is a fact that arises due to filter coefficients not being ideal as they have been approximated by small integers. Thus, the performance level is mainly set by the size of the optimization search space.

It is also noted that the proposed approximation, due to its numerical structure, would not directly work with conventional windowing functions. However, these functions can be modified to achieve the desired windowing performance.
The idea of the proposed transform is to generate multiple beams simultaneously, and would serve the applications that need simultaneous multiple look-directions with the sharpest beams.

\subsection{Fast Algorithm for Computing the 32-point ADFT}

A fast algorithm for computing the approximate
transform $\hat{\mathbf{F}}_{32}$ in \eqref{eq:adft32}
to be used in place of usual FFTs can be derived
by means of
sparse matrix factorization
in a decimation-in-frequency approach~\cite{Blahut2010}.
The matrix transform~$\hat{\mathbf{F}}_{32}$
can be factorized as follows:
\begin{align}
\label{eq-fast_algorithm}
\hat{\mathbf{F}}_{32}
=
\mathbf{W}_8
\cdot
\mathbf{W}_7
\cdot
\mathbf{W}_6
\cdot
\mathbf{W}_5
\cdot
\mathbf{W}_4
\cdot
\mathbf{W}_3
\cdot
\mathbf{W}_2
\cdot
\mathbf{W}_1
,
\end{align}
where~$\mathbf{W}_i$ for~$i \in \{1,2,3,4,5,6,7,8\}$ are
sparse matrices (factorization stages).
The non-zero matrix elements of each matrix $\mathbf{W}_i$
are given in
the Appendix.
The matrix factorization in \eqref{eq-fast_algorithm} is not unique (i.e., can admit multiple different factorizations) unlike factorization of a composite integer~\cite{burton2010elementary}.
The number of stages (i.e., sparse matrices) in the matrix factorization depend on the factorization method employed.
The number of stages is not important
as long as the overall number of elementary arithmetic operations
in the factorized form
is lower
when compared to
the direct non-factorized form of the matrix-vector product.

Notice that the entries of the sparse matrices $\mathbf{W}_i$
only contain the elements from the set $\mathbb{P}_0=\{0, +1, -1, +j, -j\}$
which imply trivial arithmetic operations.
\begin{table}[h!]
\centering
\caption{Comparison of arithmetic complexities for performing the 32-point DFT using different FFT algorithms}
\label{32pointFFTComplexities}
\begin{tabular}{lcc}
\toprule
Method & \begin{tabular}[c]{@{}c@{}}No. of \\ real additions\end{tabular} & \begin{tabular}[c]{@{}c@{}}No.  of \\ real multipliers\end{tabular} \\
\midrule
Radix-2 FFT~\cite[p.~76]{Blahut2010}          & 408 & 88 \\
Split-Radix FFT~\cite{Duhamel1984}            & 388 & 68 \\
Winograd FFT~\cite{Winograd1980}              & 388 & 68 \\
Direct Computation~$\mathbf{\hat{F}}_{32}$    & 1984 & 0 \\
Fast Algorithm~$\mathbf{\hat{F}}_{32}$        & 348 & 0 \\
\bottomrule
\end{tabular}
\end{table}
\begin{table}

\centering
\caption{Comparison of ASIC realization metrics for the proposed ADFT vs a 32-point FFT (Duhamel) using a 45-nm PDK}
\label{ASIC_comp}
\begin{tabular}{p{2.5cm} c c c}
\toprule
Metric & Duhamel algorithm & ADFT & Change
\\
\midrule
Area, A (mm$^2$) &
0.856 & 0.465 & 46\%$ \downarrow$
\\
\midrule
Critical path delay, $T$~(ns) &
1.73 & 0.86 & 50\%$\downarrow $
\\
\midrule
Frequency, $F_{max}$~(GHz) &
0.58 & 1.16 & 100\%$\uparrow$
\\
\midrule
AT ($\text{mm}^2 \cdot \text{ns}$) &
1.481 & 0.400 & 73\%$\downarrow$
\\
\midrule
AT$^2$ ($\text{mm}^2 \cdot \text{ns}^2$) &
2.562 & 0.344 & 86\%$\downarrow$
\\
\midrule
Dynamic Power, $D_p$~(mW/GHz) &
1303 & 580 & 55\%$\downarrow$
\\
\midrule
Largest side-lobe level~(dB) &
$-13.26$ & $-11.03$ & 2.23$\uparrow$
\\
\bottomrule
\end{tabular}
\flushleft \footnotesize{**The proposed algorithm achieves $\approx 50\%$ reduction in area and time at the expense of $\approx 2$dB increase in side lobe levels.}
\end{table}
Given the fast algorithm in \eqref{eq-fast_algorithm},
the computational complexity associated with computing can be quantified.
Let us consider the complex input signals which correspond to inputs being
the
I and Q outputs of the received signal from the array to the digital processor and evaluate the
arithmetic complexity in terms of real operations.
The arithmetic cost of each matrix in each factorization stage
of \eqref{eq-fast_algorithm} is evaluated
as described in~\cite{Blahut2010}.
Because the coefficients of the real and imaginary
parts of~$\mathbf{W}_i$ for~$i \in \{1,2,3,4,5,6,7,8\}$
are also in~$\mathbb{P}_0$,
only additions
are required;
multiplications and bit-shifting operations
are absent.
The additive cost is based
on the number of nonzero elements
the rows of each $\mathbf{W}_i$ matrix,
as detailed in~\cite{Blahut2010}.
Therefore,
the matrices~$\mathbf{W}_1$,~$\mathbf{W}_2$, and~$\mathbf{W}_5$
require 60~real additions;
the matrices~$\mathbf{W}_3$,~$\mathbf{W}_4$, and~$\mathbf{W}_6$
require 28~real additions;
and
the matrix~$\mathbf{W}_7$ requires 24~real additions.
The only complex matrix in the factorization,~$\mathbf{W}_8$,
requires 60~real additions only.
In total, the transform~$\hat{\mathbf{F}}_{32}$
requires~348 real additions.
Table~\ref{32pointFFTComplexities} shows
the real multiplicative and additive costs
associated with several well-known FFT algorithms
compared with the proposed algorithm.
Table~\ref{32pointFFTComplexities} also shows the additive complexity achieved through the proposed fast algorithm
is 40\% lower when compared to direct computation of $\hat{\mathbf{F}}_{32}$.

\subsection{Hardware Metrics of the Proposed ADFT Realization}

The 32-point ADFT fast algorithm in \eqref{eq-fast_algorithm} was realized as a digital core and synthesized using 45~nm CMOS free-PDK standard cells~\cite{45nmFreePDK}. For comparison purposes, a 32-point FFT core based on the Duhamel algorithm was implemented digitally and synthesized using the same technology. Both the approximate and fixed point exact FFT digital cores assume inputs of 8-bit word length. The fixed-point exact FFT core was designed with 10-bit twiddle factors \cite{FFT_twiddle_factor} which maintains a precision of $2^{-9}$ in the phasing coefficients. The multiplications throughout the signal paths were handled such that they preserve at least the coefficient precision.
Table~\ref{ASIC_comp} compares the following metrics for the two implementations:
chip area $A$,
critical path delay $T$,
maximum clock frequency $F_{max}$,
area-time $AT$,
area-time-squared $AT^2$,
frequency- and voltage-normalized dynamic power consumption $D_p$,
and
maximum side-lobe level.
It can be seen that the proposed ADFT algorithm consumes 46\% less area than the reference FFT-based design, while achieving a 50\% drop in critical path delay. It is also noted that the metrics $AT$ and $AT^2$ are reduced by $73\%$ and $86\%$, respectively, where the metric $AT$ is important when area/cost is more important, $AT^2$ is critical when speed performance is crucial. Note that the speed values mentioned in Table~\ref{ASIC_comp} are only based on synthesis results, i.e., do not consider layout effects that slow down the performance of physical implementations. However, such effects will be present in both designs, so the relative improvements in $AT$ and $AT^2$ metrics are expected to remain valid.
The compromise is an $\approx 2$dB increase in sidelobe level, which we assume to be tolerable in most RF beamforming applications where unwanted signals (jammers) can fall on larger sidelobes.

\subsection{$N$-Beam Beamforming Architectures for ULAs and URAs}
Fig.~\ref{fig_ULA_URA_archs} shows the top-level hardware architectures for realizing $N$ and $N^2$ simultaneous orthogonal beams for an $N$-element and $N \times N$ aperture respectively using an $N$-point spatial DFT digital core as the basic signal processing block. The front-end is shown as a direct-conversion receiver chain followed by analog-to-digital conversion for digital beamforming. The digitized data can be converted to complex ($I$-$Q$) form using a Hilbert transform. This can also be achieved by using a quadrature mixer in the RF chain, with the luxury of going to baseband directly at the cost of double the amount of ADCs.
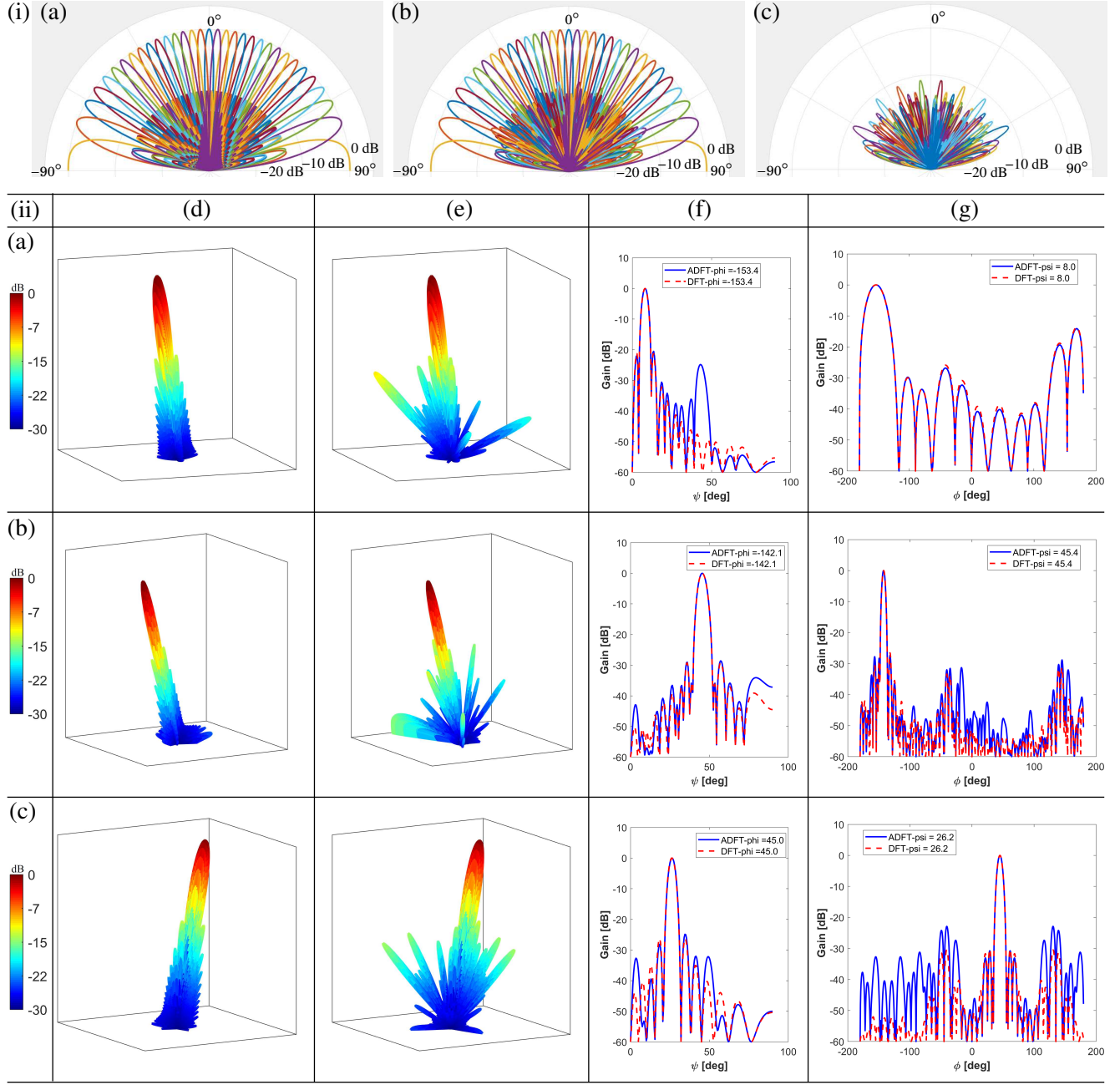
\begin{figure*}

\centering
\scalebox{1.1}{\input{ULA_URA_sim_beams_R1.pstex_t}}
\caption{(i) Simulated polar patterns of the 32-beams for a ULA with $\lambda/2$ element spacing. (i-a) Beams corresponding to the ADFT, (i-b) beams obtained
with the ideal FFT, and (i-c) the magnitude error between the ADFT and the exact FFT.
(ii) Example simulated beam patterns from a Nyquist-spaced URA; (a) $\psi=8.0^{\circ},\phi=-153.4^{\circ}$, (b) $\psi=45.4^{\circ},\phi= -142.1^{\circ}$,(c) $\psi=26.2^{\circ},\phi=45.0^{\circ}$ (the plots are color-coded on a dB scale).\label{fig_ula_ura_sim_beam}}
\vskip -2ex
\end{figure*}

The numerically simulated array factors resulting from a 32-element spatially Nyquist sampled ULA are given in Fig.~\ref{fig_ula_ura_sim_beam}(i). Fig.~\ref{fig_ula_ura_sim_beam}(i-b) shows the beams generated using the proposed 32-point ADFT algorithm and Fig.~\ref{fig_ula_ura_sim_beam}(i-a) shows the corresponding beams of the exact algorthm with (i-c) showing the error magnitude between them.
Fig.~\ref{fig_ula_ura_sim_beam}(ii) shows three simulated example beams out of the 1024 beams generated by the proposed ADFT algorithm when it is applied to a $32\times 32$-element URA.
The first and second columns of Fig.~\ref{fig_ula_ura_sim_beam}(ii) show the beams corresponding to the exact and approximate DFT, respectively; the third and fourth columns of Fig.~\ref{fig_ula_ura_sim_beam}(ii) show the errors between the two algorithms in the elevation and azimuthal planes, respectively, which are small enough to be ignored for most microwave and mm-wave beamforming applications.

\section{A 32-Beam ULA-based Multibeam Beamformer}

The system architecture used for verifying the proposed low-complexity multibeam beamforming algorithm is shown in Fig.~\ref{rf_setup}(a).
This section explains the system design.

\subsection{5.8~GHz Front-End Design}
\begin{figure*}
\centering
\scalebox{1.0}{\input{overview_rf.pstex_t}}
\caption{(a) Overall architecture of the test setup; (b) \href{https://ieeetv.ieee.org/ieeetv-specials/ted-tours-the-brooklyn-5g-summit-expo-floor-3}{5.8~GHz 32-beam array receiver setup} (starting at 2m:44s in the hyperlinked video); (c) measured S-parameters (return loss and mutual coupling) of the designed antenna array.}
\label{rf_setup}
\end{figure*}
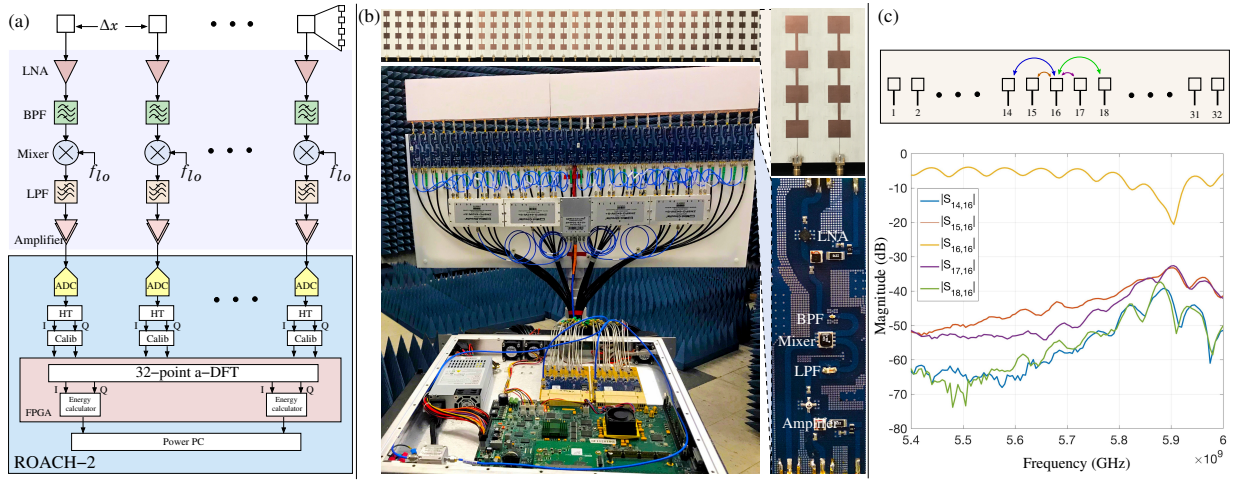

\begin{table}
\centering
\caption{Design Specifications for the Patch Antenna}
\label{design_specs}
\begin{tabular}{lc}
\toprule
Frequency $(f_0)$      & 5.8~GHz \\
\midrule
Substrate              & Rogers RO4350B \\
\midrule
Dielectric constant $\left( \epsilon _r \right)$ & 3.66 \\
\midrule
Substrate thickness ($h$)                        & 0.508~mm \\
\midrule
Conductor (copper) thickness ($ct$)              & 35~$\mu$m  \\
\bottomrule
\end{tabular}
\end{table}
The RF front-end of the receive-mode beamformer is constructed by integrating a 32-element ULA at 5.8~GHz with 32 direct conversion RF receiver chains (on PCB) as shown in Fig.~\ref{rf_setup}(b). The inter-element spacing of the array $\Delta x$ was set to $0.6\lambda$, which is $\approx 31$~mm at 5.8~GHz.

Each antenna element of the ULA was designed as a $4 \times 1$ vertical sub-array of patch antennas that employs passive beamforming at RF in the orthogonal (vertical) plane. This design improves the gain in the vertical plane, thus simplifying array factor measurements in the azimuthal plane. In principle, such analog beamforming is independent of the 1D/2D spatial FFT-based beamforming algorithms discussed in this paper.
The sub-array is designed by feeding antenna elements in series along a uniform transmission line, and performing a parametric sweep to provide better impedance matching and performance~\cite{sainati}. Note that such analog beamforming does not affect the performance of the beamforming algorithm under consideration as it happens in the azimuthal plane. The specifications of the antenna array design are summarized in Table~\ref{design_specs}.
The antenna outputs are directly fed into 32 heterodyne receivers designed on FR-4 PCBs using surface mount devices. The LO signals for each receiver are provided through a centralized LO scheme that consists of a 32-output power divider network connected to a low-phase-noise  oscillator.  The first stage of each receiver consists of a low-noise amplifier (LNA) that provides 16~dB gain at 5.8~GHz with a noise figure of 2.4~dB. The amplified signal is band-pass filtered within the frequency range 4.7-6 GHz, which helps to reject out-of-band interference and noise. The band-limited amplified signal is then passed through a mixer and low-pass filter to produce a downconverted low-IF input. The 32 downconverted low-IF signals are further amplified by $\sim$30~dB and then digitized in parallel using two ADC16x250-8 ADC cards (16 single-ended input channels, 8-bit, up to 250~MS/s per channel)~\cite{ADC16x250-8}. The in-band gain and noise figure of the entire receiver are estimated to be 38.6~dB and 2.9~dB, respectively; the latter is dominated by the LNA.

The ADCs used a sample clock of 200~MHz for real-time hardware experiments. The same clock was also routed to the digital circuits implemented on the FPGAs. The clock frequency was chosen to be smaller than the maximum allowed for the digital design, which is limited by the critical path delay (CPD) and thus denoted by $f_{max,CPD}$.

\subsection{Digital Back-End}
Digital processing of sampled signals was performed using the reconfigurable open architecture computing hardware (\mbox{ROACH-2}) platform~\cite{roach_2} designed by the Collaboration for Astronomy Signal Processing and Electronics Research (CASPER). ROACH-2 is based on a Xilinx Virtex-6 FPGA chip; it also includes an integrated on-board processor that handles communications and control functions with the FPGA.
The platform has 2 ZDOK interfaces (each supporting 42 pins) that connects high bandwidth input/output (I/O) to the FPGA.  The ADC16x250-8 ADC cards mentioned previously were designed to be compatible with the ROACH-2 hardware and can be accessed using CASPER-supplied software routines. These routines, which are available at~\cite{casper-adc-cal}, also allow the ADCs to be calibrated.

The overall architecture of the digital beamforming test-setup is shown in Fig.~\ref{rf_setup}(a). The digital design consists of four main subsystems: (i) a digital calibration stage; (ii) an IQ decomposition FIR filter that implements the Hilbert transform~\cite{Oppenheim2009}; (iii) the 32-point DFT/ADFT algorithm implementation; and (iv) an energy calculation subsystem for facilitating real-time measurements on each output beam. The exact DFT core was designed using 10-bit precision twiddle factors which provide a good compromise between circuit size and maximum operating frequency.

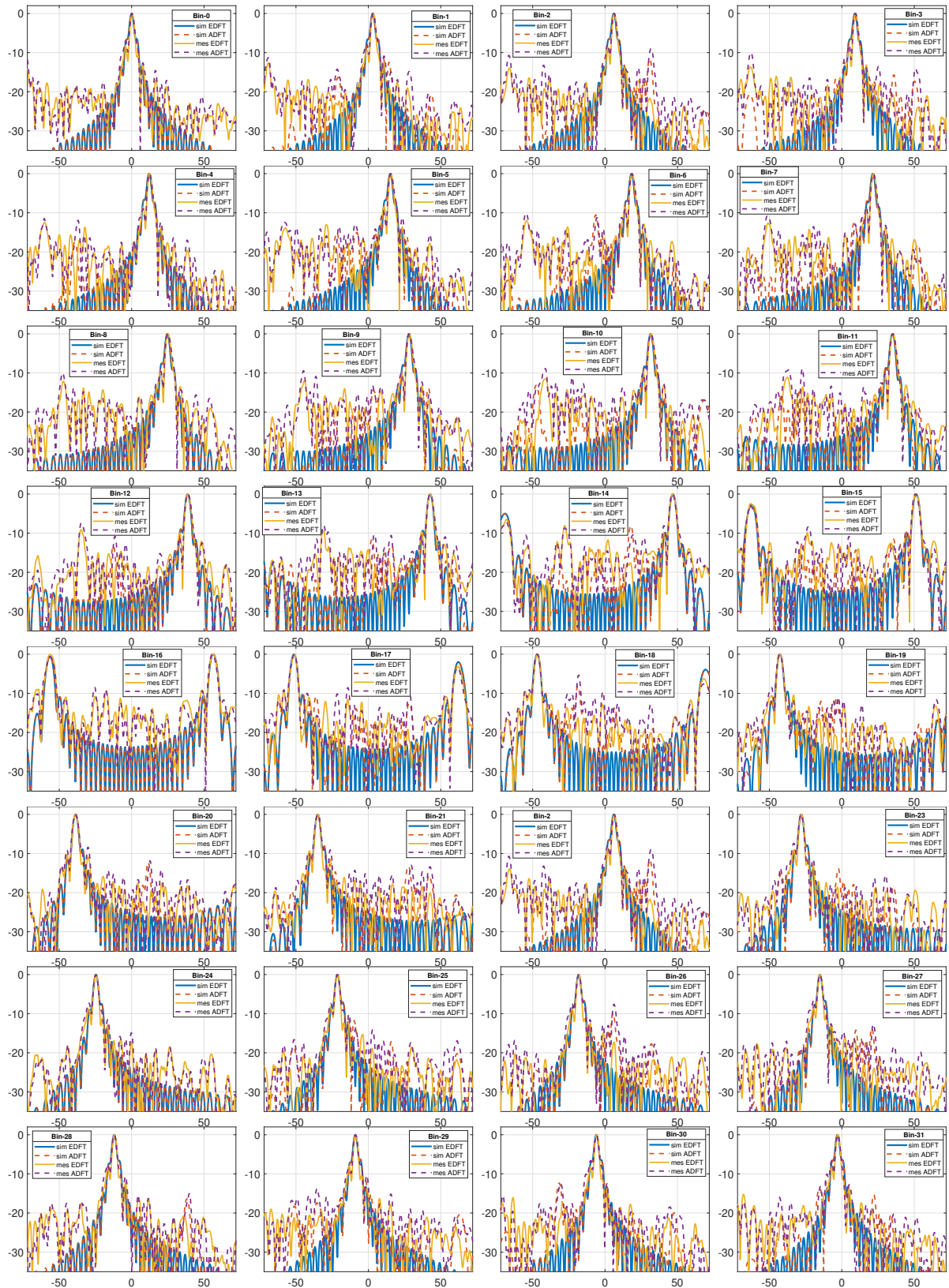
\begin{figure*}
	\centering
	\scalebox{0.81}{\input{mes_beams.pstex_t}}
	\caption{The 32 beams measured from the 5.8~GHz array (vertical axis is in dB and the horizontal axis is the azimuthal angle $[-72,72]$). Each sub figure contains the measured beams pattern using using the fixed-point FFT digital core, the ADFT core and the simulated beams for both scenarios. Simulated beams accounts antenna element pattern and the actual separation of the transmitter and the receiver in the measurement setup. \label{fig_mesResults}}
\end{figure*}

\section{Experimental Results}
This section describes experimental results obtained from the 32-element ULA, including antenna characterization and beam measurements. %

\subsection{Antenna Array Characterization}
The performance of the array can be characterized using S-parameters~\cite{pozar}, which were measured using a commercial vector network analyzer. For example, the return loss of the $m$th antenna, which represents the amount of power reflected from it, is given by $|S_{m,m}|$. The measured return loss for $m=16$, namely $|S_{16,16}|$, is shown in Fig.~\ref{rf_setup}(c). The proposed patch antenna resonates at a frequency of 5.9~GHz with an excellent return loss of $-20.6$~dB.

Mutual coupling is another important issue during the design of an antenna array. In an array, the fields radiated by individual elements tend to interact with each other, thus causing interchange of energy~\cite{mc_2}. Mutual coupling describes the energy absorbed by one antenna when a nearby antenna is operating, and depends upon many factors including antenna design parameters, inter-element spacing, and the direction of arrival (DOA) of the wave~\cite{mc_1}. It can also be measured using S-parameters. Specifically, $S_{n,m}$ is a measure of coupling between ports $m$ and $n$. We measured the mutual coupling between an antenna element and its nearest neighbors using $S_{n,m}$ measurements. In particular, we measured $S_{14,16}$, $S_{15,16}$, $S_{17,16}$ and $S_{18,16}$, which characterize the coupling between ports 16 and its near neighbors (ports 14, 15, 17, and 18). The results are shown in Fig.~\ref{rf_setup}(c) versus frequency: the values at 5.8~GHz are relatively low and given by
$|S_{14,16}| = -39.2$~dB,
$|S_{15,16}| = -33.2$~dB,
$|S_{17,16}| = -33.0$~dB,
and
$|S_{18,16}|=-37.3$~dB.
As expected, mutual coupling decreases with inter-element separation.

\subsection{Calibration}
Calibration of the RF array system is vital for obtaining optimal beamforming performance. Calibration was performed in two stages. The first stage was performed on the ADCs, and used open source routines that have already been developed for the same hardware by members of the CASPER group~\cite{casper-adc-cal}.
The second stage focused on digitally removing the effects of mismatches in the microwave front-end. Relative gain and phase mismatches of the IF outputs for each chain were calculated with respect to a reference chain using a input reference carrier at 5.86~GHz. Since the overall system is narrowband, the recorded gain and phase values were directly used to equalize the gain and phase of the sampled IF inputs. This was achieved by adding a complex multiplier after the digital Hilbert transform in each channel.

\begin{figure}[h!]
\centering
\scalebox{0.85}{\input{Fig_separation.pstex_t}}
\vskip -2ex
\caption{The impact of the measurement setup geometry on the measured beam response. \label{Fig_separation}}
\vskip -2ex
\end{figure}
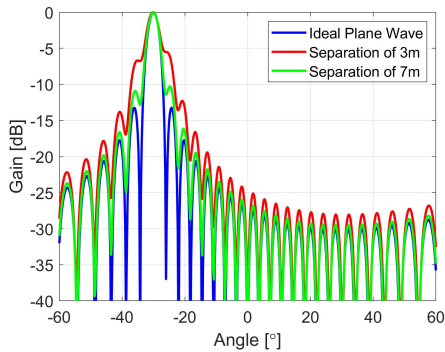

\subsection{Beam Measurements}
As shown in Fig.~\ref{rf_setup}(b), the entire 5.8~GHz 32-element digital array placed in an anechoic chamber for measuring the received beam patterns (\cite{B5GSummitVideo} shows a short realtime demo of the total system). Power patterns were measured by sending a continuous wave (CW) signal at $f_{RF} = 5.86$~GHz. The LO signal frequency $f_{LO}$ determines the IF ${f_{RF} - f_{LO}}$. The measurements in this paper were generated by setting $f_{LO} = 5.85$~GHz, thus resulting in an IF of 10~MHz, and digitizing the down-converted outputs at $f_{clk}=200$~MHz. Fig.~\ref{fig_mesResults} shows the measured beams from the real-time experimental setup for both the exact and approximate algorithms along with the corresponding simulated curves. The measurement was conducted by using digital integrators at each FFT/ADFT bin output to calculate the received energy for a fixed amount of time.

The measured array factor of the beams highly depends on the measurement setup geometry. Ideally, the transmitter and the receiver should be placed far enough apart for waves incident on the receiver array to be approximated as plane waves. Numerical simulation in Fig.~\ref{Fig_separation} shows how the actual array factor being measured deviates from this ideal depending on the geometry of the test setup. Based on the standard rules \cite[p. 42]{AntennaTheoryDesign3rdEd}, the transmitter and receiver should have a separation exceeding  20 m at 5.8~GHz in order for the receiver aperture to be in the far field. However, such a large separation was not achievable within our test facility. In particular, the beams were measured in an open parking deck with a transmitter receiver separation of approximately 7~m. Due to this reason, the measured beams shown in Fig.~\ref{fig_mesResults} have been compared with numerically-simulated beams that account for both finite transmitter-receiver separation and the actual element pattern.

\begin{figure*}[t!]
	\centering
	\scalebox{1.3}{\input{mes_sim_2D_beams.pstex_t}}
	\vskip -3ex
	\caption{2D beam patterns computed from 1D array beam measurements using the ADFT algorithm. The beams correspond to the bin outputs (same angles) as the beams shown in Fig.~\ref{fig_ula_ura_sim_beam}(a-c). \label{fig_mes_sim_beams}}
	\vskip -2ex
\end{figure*}
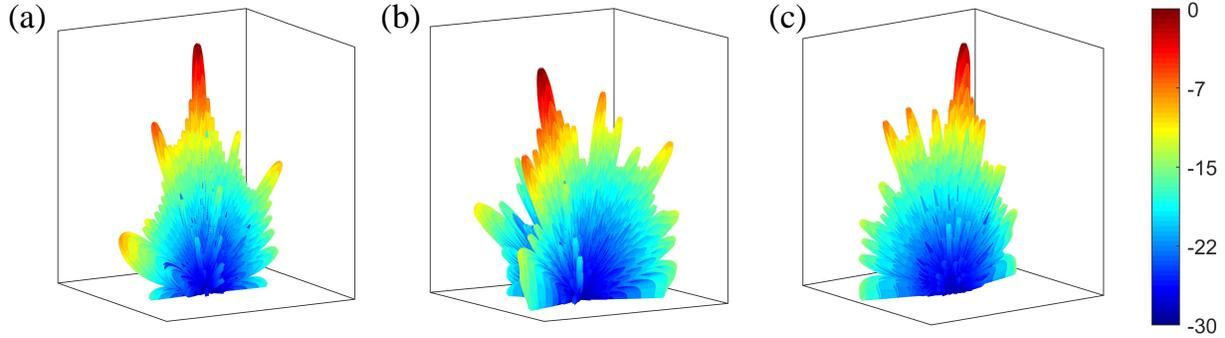

Fig.~\ref{fig_mesResults} shows that  the measured beam patterns for both the algorithms closely follow each other for all the bins. The measured beams also follow the expected patterns quite well in the vicinity of the main beam. For both algorithms, the measured plots have higher side-lobe levels in the deeper stop bands compared to the simulated ones. We believe that such degradation in stop band performance is mainly due to post-calibration errors of the system; these are dominated by the performance of the analog front-ends in the receiver. Measurement errors, including the fact that the tests were not performed in an anechoic environment, also lead to deviations from the expected patterns.

The 2D array factor of each beam arising from the proposed linear transform can be expressed as,
\begin{align}
\Psi_{k,l}(\omega_x,\omega_y) = \sum_{n=0}^{31}\sum_{m=0}^{31}
\left[
\mathbf{\hat{F}}_{32}(k,m)
\mathbf{\hat{F}}_{32}(l,n)
e^{-j(\omega_x \Delta x m +\omega_y \Delta y n)}
\right]
,
\label{eq-2D_Beams_1}
\end{align}
which may be rearranged to
\begin{align}
\Psi_{k,l}(\omega_x,\omega_y)
&=
\left(
\sum_{m=0}^{31}  \mathbf{\hat{F}}_{32}(k,m) e^{-j\omega_x \Delta x m}
\right)
\times
\left(
\sum_{n=0}^{31} \mathbf{\hat{F}}_{32}(l,n) e^{-j\omega_y \Delta y n}
\right) \\
&=
\Upsilon(k,\omega_x)\times \sum_{n=0}^{31}\mathbf{\hat{F}}_{32}(l,n) e^{-j\omega_y \Delta y n}
,
\label{eq-2D_Beams_2}
\end{align}
where $k,l \in [0,1,\ldots,31]$, $\omega_x = \omega_{ct} \sin \psi\cos \phi$, $\omega_y = \omega_{ct} \sin \psi\cos \phi$, $\psi$ and $\phi$ are elevation and azimuthal angles, respectively. $\Delta x$ and $\Delta y$ denote the inter-element spacing in $x$ and $y$ directions.
The relationship in (\ref{eq-2D_Beams_2}) can be used to compute the 2D beam responses corresponding to a 2D URA consisting of 32 linear arrays, each with the measured responses shown in Fig.~\ref{fig_mesResults}. In particular, the term $\Upsilon(k,\omega_x)$ denotes the array factor of the $k$th beam in the 32-element linear array subsystem. The measured 1D beam patterns were thus used in place of $\Upsilon(k,\omega_x)$ to synthesize the corresponding 2D beam patterns from a 2D aperture.
Fig.~\ref{fig_mes_sim_beams} shows the 2D beam patterns obtained using the measured ULA beam measurements for the same beams shown in Fig.~\ref{fig_ula_ura_sim_beam} assuming $\Delta x = \Delta y$.

\section{Conclusion}
A large number of simultaneous beams has become an essential requirement for emerging mm-wave based 5G systems. Moreover, future communications applications, such as space-based Internet services, demand an ultra-high number of beams.
An $N \times N$ square antenna array aperture can generate up to $N^2$ orthogonal simultaneous beams by using the 2D $N$-point spatial DFT.
The upper bound of the multiplicative complexity associated with such processing using FFT algorithms is $\mathcal{O}(2N^2\:\log N)$.
This paper has discussed a low-complexity digital beamforming architecture for generating 1024~simultaneous RF beams using a 32-point DFT approximation that completely eliminates multiplication operations.
The proposed ADFT algorithm consumes 46\% less
area than the reference FFT-based design, while achieving a
50\% drop in critical path delay.
The VLSI metrics $AT$ and $AT^2$ for the proposed algorithm are reduced by 73\% and 86\%, respectively.
We have validated the proposed approach on a fully-functional 32-element digital 1D receive array that operates at 5.8~GHz. This design will serve as the main subsystem for future implementations of a $32 \times 32$ 2D rectangular aperture that could generate 1024~simultaneous RF beams with significantly lower SWaP in VLSI implementations.
The 1D array uses 32~parallel ADCs for sampling the antenna outputs and the ADFT (implemented on a Xilinx FPGA) for computing 32 RF beams in real-time. The measured RF beams show a per-beam bandwidth of 120~MHz when all 32~beams are realized in real time, with only marginal ($<2$~dB) degradation in beam performance compared to a control experiment based on athe Duhamel FFT core.

\section*{Acknowledgment}
The authors are grateful to the CASPER community for thoughtful discussions and advice through their e-mail discussion group. This work would not have been possible without the extensive contributions to the open source ROACH-2 system development by multiple contributors in the radio-astronomy instrumentation community.
One of the authors (RJC) thanks an anonymous reviewer
and Mr{.} L{.} Portella for the identification of a typo
in the matrix equations from Section~3.2.

\appendix

\section{Factorization Terms of $\hat{\mathbf{F}}_{32}$}

The matrix factors $\mathbf{W}_i$, $i=1,2,\ldots,8$,
from \eqref{eq-fast_algorithm}
are given below:
\begin{align}
\mathbf{W}_1
=
\begin{bsmallmatrix*}[r]
\phantom{-}1 & \phantom{-}0 & \phantom{-}0 & \phantom{-}0 & \phantom{-}0 & \phantom{-}0 & \phantom{-}0 & \phantom{-}0 & \phantom{-}0 & \phantom{-}0 & \phantom{-}0 & \phantom{-}0 & \phantom{-}0 & \phantom{-}0 & \phantom{-}0 & \phantom{-}0 & \phantom{-}1 & \phantom{-}0 & \phantom{-}0 & \phantom{-}0 & \phantom{-}0 & \phantom{-}0 & \phantom{-}0 & \phantom{-}0 & \phantom{-}0 & \phantom{-}0 & \phantom{-}0 & \phantom{-}0 & \phantom{-}0 & \phantom{-}0 & \phantom{-}0 & \phantom{-}0 \\
0 & 1 & 0 & 0 & 0 & 0 & 0 & 0 & 0 & 0 & 0 & 0 & 0 & 0 & 0 & 1 & 0 & 0 & 0 & 0 & 0 & 0 & 0 & 0 & 0 & 0& 0 & 0 & 0 & 0 & 0 & 0 \\
0 & 0 & 1 & 0 & 0 & 0 & 0 & 0 & 0 & 0 & 0 & 0 & 0 & 0 & 1 & 0 & 0 & 0 & 0 & 0 & 0 & 0 & 0 & 0 & 0 & 0& 0 & 0 & 0 & 0 & 0 & 0 \\
0 & 0 & 0 & 1 & 0 & 0 & 0 & 0 & 0 & 0 & 0 & 0 & 0 & 1 & 0 & 0 & 0 & 0 & 0 & 0 & 0 & 0 & 0 & 0 & 0 & 0& 0 & 0 & 0 & 0 & 0 & 0 \\
0 & 0 & 0 & 0 & 1 & 0 & 0 & 0 & 0 & 0 & 0 & 0 & 1 & 0 & 0 & 0 & 0 & 0 & 0 & 0 & 0 & 0 & 0 & 0 & 0 & 0& 0 & 0 & 0 & 0 & 0 & 0 \\
0 & 0 & 0 & 0 & 0 & 1 & 0 & 0 & 0 & 0 & 0 & 1 & 0 & 0 & 0 & 0 & 0 & 0 & 0 & 0 & 0 & 0 & 0 & 0 & 0 & 0& 0 & 0 & 0 & 0 & 0 & 0 \\
0 & 0 & 0 & 0 & 0 & 0 & 1 & 0 & 0 & 0 & 1 & 0 & 0 & 0 & 0 & 0 & 0 & 0 & 0 & 0 & 0 & 0 & 0 & 0 & 0 & 0& 0 & 0 & 0 & 0 & 0 & 0 \\
0 & 0 & 0 & 0 & 0 & 0 & 0 & 1 & 0 & 1 & 0 & 0 & 0 & 0 & 0 & 0 & 0 & 0 & 0 & 0 & 0 & 0 & 0 & 0 & 0 & 0& 0 & 0 & 0 & 0 & 0 & 0 \\
0 & 0 & 0 & 0 & 0 & 0 & 0 & 0 & 1 & 0 & 0 & 0 & 0 & 0 & 0 & 0 & 0 & 0 & 0 & 0 & 0 & 0 & 0 & 0 & 0 & 0& 0 & 0 & 0 & 0 & 0 & 0 \\
0 & 0 & 0 & 0 & 0 & 0 & 0 & 1 & 0 & -1 & 0 & 0 & 0 & 0 & 0 & 0 & 0 & 0 & 0 & 0 & 0 & 0 & 0 & 0 & 0 & 0 & 0 & 0 & 0 & 0 & 0 & 0 \\
0 & 0 & 0 & 0 & 0 & 0 & 1 & 0 & 0 & 0 & -1 & 0 & 0 & 0 & 0 & 0 & 0 & 0 & 0 & 0 & 0 & 0 & 0 & 0 & 0 & 0 & 0 & 0 & 0 & 0 & 0 & 0 \\
0 & 0 & 0 & 0 & 0 & 1 & 0 & 0 & 0 & 0 & 0 & -1 & 0 & 0 & 0 & 0 & 0 & 0 & 0 & 0 & 0 & 0 & 0 & 0 & 0 & 0 & 0 & 0 & 0 & 0 & 0 & 0 \\
0 & 0 & 0 & 0 & 1 & 0 & 0 & 0 & 0 & 0 & 0 & 0 & -1 & 0 & 0 & 0 & 0 & 0 & 0 & 0 & 0 & 0 & 0 & 0 & 0 & 0 & 0 & 0 & 0 & 0 & 0 & 0 \\
0 & 0 & 0 & 1 & 0 & 0 & 0 & 0 & 0 & 0 & 0 & 0 & 0 & -1 & 0 & 0 & 0 & 0 & 0 & 0 & 0 & 0 & 0 & 0 & 0 & 0 & 0 & 0 & 0 & 0 & 0 & 0 \\
0 & 0 & 1 & 0 & 0 & 0 & 0 & 0 & 0 & 0 & 0 & 0 & 0 & 0 & -1 & 0 & 0 & 0 & 0 & 0 & 0 & 0 & 0 & 0 & 0 & 0 & 0 & 0 & 0 & 0 & 0 & 0 \\
0 & 1 & 0 & 0 & 0 & 0 & 0 & 0 & 0 & 0 & 0 & 0 & 0 & 0 & 0 & -1 & 0 & 0 & 0 & 0 & 0 & 0 & 0 & 0 & 0 & 0 & 0 & 0 & 0 & 0 & 0 & 0 \\
1 & 0 & 0 & 0 & 0 & 0 & 0 & 0 & 0 & 0 & 0 & 0 & 0 & 0 & 0 & 0 & -1 & 0 & 0 & 0 & 0 & 0 & 0 & 0 & 0 & 0 & 0 & 0 & 0 & 0 & 0 & 0 \\
0 & 0 & 0 & 0 & 0 & 0 & 0 & 0 & 0 & 0 & 0 & 0 & 0 & 0 & 0 & 0 & 0 & 1 & 0 & 0 & 0 & 0 & 0 & 0 & 0 & 0& 0 & 0 & 0 & 0 & 0 & 1 \\
0 & 0 & 0 & 0 & 0 & 0 & 0 & 0 & 0 & 0 & 0 & 0 & 0 & 0 & 0 & 0 & 0 & 0 & 1 & 0 & 0 & 0 & 0 & 0 & 0 & 0& 0 & 0 & 0 & 0 & 1 & 0 \\
0 & 0 & 0 & 0 & 0 & 0 & 0 & 0 & 0 & 0 & 0 & 0 & 0 & 0 & 0 & 0 & 0 & 0 & 0 & 1 & 0 & 0 & 0 & 0 & 0 & 0& 0 & 0 & 0 & 1 & 0 & 0 \\
0 & 0 & 0 & 0 & 0 & 0 & 0 & 0 & 0 & 0 & 0 & 0 & 0 & 0 & 0 & 0 & 0 & 0 & 0 & 0 & 1 & 0 & 0 & 0 & 0 & 0& 0 & 0 & 1 & 0 & 0 & 0 \\
0 & 0 & 0 & 0 & 0 & 0 & 0 & 0 & 0 & 0 & 0 & 0 & 0 & 0 & 0 & 0 & 0 & 0 & 0 & 0 & 0 & 1 & 0 & 0 & 0 & 0& 0 & 1 & 0 & 0 & 0 & 0 \\
0 & 0 & 0 & 0 & 0 & 0 & 0 & 0 & 0 & 0 & 0 & 0 & 0 & 0 & 0 & 0 & 0 & 0 & 0 & 0 & 0 & 0 & 1 & 0 & 0 & 0& 1 & 0 & 0 & 0 & 0 & 0 \\
0 & 0 & 0 & 0 & 0 & 0 & 0 & 0 & 0 & 0 & 0 & 0 & 0 & 0 & 0 & 0 & 0 & 0 & 0 & 0 & 0 & 0 & 0 & 1 & 0 & 1& 0 & 0 & 0 & 0 & 0 & 0 \\
0 & 0 & 0 & 0 & 0 & 0 & 0 & 0 & 0 & 0 & 0 & 0 & 0 & 0 & 0 & 0 & 0 & 0 & 0 & 0 & 0 & 0 & 0 & 0 & 1 & 0& 0 & 0 & 0 & 0 & 0 & 0 \\
0 & 0 & 0 & 0 & 0 & 0 & 0 & 0 & 0 & 0 & 0 & 0 & 0 & 0 & 0 & 0 & 0 & 0 & 0 & 0 & 0 & 0 & 0 & 1 & 0 & -1 & 0 & 0 & 0 & 0 & 0 & 0 \\
0 & 0 & 0 & 0 & 0 & 0 & 0 & 0 & 0 & 0 & 0 & 0 & 0 & 0 & 0 & 0 & 0 & 0 & 0 & 0 & 0 & 0 & 1 & 0 & 0 & 0& -1 & 0 & 0 & 0 & 0 & 0 \\
0 & 0 & 0 & 0 & 0 & 0 & 0 & 0 & 0 & 0 & 0 & 0 & 0 & 0 & 0 & 0 & 0 & 0 & 0 & 0 & 0 & 1 & 0 & 0 & 0 & 0& 0 & -1 & 0 & 0 & 0 & 0 \\
0 & 0 & 0 & 0 & 0 & 0 & 0 & 0 & 0 & 0 & 0 & 0 & 0 & 0 & 0 & 0 & 0 & 0 & 0 & 0 & 1 & 0 & 0 & 0 & 0 & 0& 0 & 0 & -1 & 0 & 0 & 0 \\
0 & 0 & 0 & 0 & 0 & 0 & 0 & 0 & 0 & 0 & 0 & 0 & 0 & 0 & 0 & 0 & 0 & 0 & 0 & 1 & 0 & 0 & 0 & 0 & 0 & 0& 0 & 0 & 0 & -1 & 0 & 0 \\
0 & 0 & 0 & 0 & 0 & 0 & 0 & 0 & 0 & 0 & 0 & 0 & 0 & 0 & 0 & 0 & 0 & 0 & 1 & 0 & 0 & 0 & 0 & 0 & 0 & 0& 0 & 0 & 0 & 0 & -1 & 0 \\
0 & 0 & 0 & 0 & 0 & 0 & 0 & 0 & 0 & 0 & 0 & 0 & 0 & 0 & 0 & 0 & 0 & 1 & 0 & 0 & 0 & 0 & 0 & 0 & 0 & 0& 0 & 0 & 0 & 0 & 0 & -1 \\
\end{bsmallmatrix*}
,
\end{align}

\begin{align}
\mathbf{W}_2
=
\begin{bsmallmatrix*}[r]
\phantom{-}1 & \phantom{-}0 & \phantom{-}0 & \phantom{-}0 & \phantom{-}0 & \phantom{-}0 & \phantom{-}0 & \phantom{-}0 & \phantom{-}0 & \phantom{-}0 & \phantom{-}0 & \phantom{-}0 & \phantom{-}0 & \phantom{-}0 & \phantom{-}0 & \phantom{-}0 & \phantom{-}0 & \phantom{-}0 & \phantom{-}0 & \phantom{-}0 & \phantom{-}0 & \phantom{-}0 & \phantom{-}0 & \phantom{-}0 & \phantom{-}0 & \phantom{-}0 & \phantom{-}0 & \phantom{-}0 & \phantom{-}0 & \phantom{-}0 & \phantom{-}0 & \phantom{-}0 \\
0 & 1 & 0 & 0 & 0 & 0 & 0 & 0 & 0 & 0 & 0 & 0 & 0 & 0 & 0 & 0 & 0 & 1 & 0 & 0 & 0 & 0 & 0 & 0 & 0 & 0& 0 & 0 & 0 & 0 & 0 & 0 \\
0 & 0 & 1 & 0 & 0 & 0 & 0 & 0 & 0 & 0 & 0 & 0 & 0 & 0 & 0 & 0 & 0 & 0 & 1 & 0 & 0 & 0 & 0 & 0 & 0 & 0& 0 & 0 & 0 & 0 & 0 & 0 \\
0 & 0 & 0 & 1 & 0 & 0 & 0 & 0 & 0 & 0 & 0 & 0 & 0 & 0 & 0 & 0 & 0 & 0 & 0 & 1 & 0 & 0 & 0 & 0 & 0 & 0& 0 & 0 & 0 & 0 & 0 & 0 \\
0 & 0 & 0 & 0 & 1 & 0 & 0 & 0 & 0 & 0 & 0 & 0 & 0 & 0 & 0 & 0 & 0 & 0 & 0 & 0 & 1 & 0 & 0 & 0 & 0 & 0& 0 & 0 & 0 & 0 & 0 & 0 \\
0 & 0 & 0 & 0 & 0 & 1 & 0 & 0 & 0 & 0 & 0 & 0 & 0 & 0 & 0 & 0 & 0 & 0 & 0 & 0 & 0 & 1 & 0 & 0 & 0 & 0& 0 & 0 & 0 & 0 & 0 & 0 \\
0 & 0 & 0 & 0 & 0 & 0 & 1 & 0 & 0 & 0 & 0 & 0 & 0 & 0 & 0 & 0 & 0 & 0 & 0 & 0 & 0 & 0 & 1 & 0 & 0 & 0& 0 & 0 & 0 & 0 & 0 & 0 \\
0 & 0 & 0 & 0 & 0 & 0 & 0 & 1 & 0 & 0 & 0 & 0 & 0 & 0 & 0 & 0 & 0 & 0 & 0 & 0 & 0 & 0 & 0 & 1 & 0 & 0& 0 & 0 & 0 & 0 & 0 & 0 \\
0 & 0 & 0 & 0 & 0 & 0 & 0 & 0 & 1 & 0 & 0 & 0 & 0 & 0 & 0 & 0 & 0 & 0 & 0 & 0 & 0 & 0 & 0 & 0 & 1 & 0& 0 & 0 & 0 & 0 & 0 & 0 \\
0 & 0 & 0 & 0 & 0 & 0 & 0 & 0 & 0 & 1 & 0 & 0 & 0 & 0 & 0 & 0 & 0 & 0 & 0 & 0 & 0 & 0 & 0 & 0 & 0 & 1& 0 & 0 & 0 & 0 & 0 & 0 \\
0 & 0 & 0 & 0 & 0 & 0 & 0 & 0 & 0 & 0 & 1 & 0 & 0 & 0 & 0 & 0 & 0 & 0 & 0 & 0 & 0 & 0 & 0 & 0 & 0 & 0& 1 & 0 & 0 & 0 & 0 & 0 \\
0 & 0 & 0 & 0 & 0 & 0 & 0 & 0 & 0 & 0 & 0 & 1 & 0 & 0 & 0 & 0 & 0 & 0 & 0 & 0 & 0 & 0 & 0 & 0 & 0 & 0& 0 & 1 & 0 & 0 & 0 & 0 \\
0 & 0 & 0 & 0 & 0 & 0 & 0 & 0 & 0 & 0 & 0 & 0 & 1 & 0 & 0 & 0 & 0 & 0 & 0 & 0 & 0 & 0 & 0 & 0 & 0 & 0& 0 & 0 & 1 & 0 & 0 & 0 \\
0 & 0 & 0 & 0 & 0 & 0 & 0 & 0 & 0 & 0 & 0 & 0 & 0 & 1 & 0 & 0 & 0 & 0 & 0 & 0 & 0 & 0 & 0 & 0 & 0 & 0& 0 & 0 & 0 & 1 & 0 & 0 \\
0 & 0 & 0 & 0 & 0 & 0 & 0 & 0 & 0 & 0 & 0 & 0 & 0 & 0 & 1 & 0 & 0 & 0 & 0 & 0 & 0 & 0 & 0 & 0 & 0 & 0& 0 & 0 & 0 & 0 & 1 & 0 \\
0 & 0 & 0 & 0 & 0 & 0 & 0 & 0 & 0 & 0 & 0 & 0 & 0 & 0 & 0 & 1 & 0 & 0 & 0 & 0 & 0 & 0 & 0 & 0 & 0 & 0& 0 & 0 & 0 & 0 & 0 & 1 \\
0 & 0 & 0 & 0 & 0 & 0 & 0 & 0 & 0 & 0 & 0 & 0 & 0 & 0 & 0 & 0 & 1 & 0 & 0 & 0 & 0 & 0 & 0 & 0 & 0 & 0& 0 & 0 & 0 & 0 & 0 & 0 \\
0 & 1 & 0 & 0 & 0 & 0 & 0 & 0 & 0 & 0 & 0 & 0 & 0 & 0 & 0 & 0 & 0 & -1 & 0 & 0 & 0 & 0 & 0 & 0 & 0 & 0 & 0 & 0 & 0 & 0 & 0 & 0 \\
0 & 0 & 1 & 0 & 0 & 0 & 0 & 0 & 0 & 0 & 0 & 0 & 0 & 0 & 0 & 0 & 0 & 0 & -1 & 0 & 0 & 0 & 0 & 0 & 0 & 0 & 0 & 0 & 0 & 0 & 0 & 0 \\
0 & 0 & 0 & 1 & 0 & 0 & 0 & 0 & 0 & 0 & 0 & 0 & 0 & 0 & 0 & 0 & 0 & 0 & 0 & -1 & 0 & 0 & 0 & 0 & 0 & 0 & 0 & 0 & 0 & 0 & 0 & 0 \\
0 & 0 & 0 & 0 & 1 & 0 & 0 & 0 & 0 & 0 & 0 & 0 & 0 & 0 & 0 & 0 & 0 & 0 & 0 & 0 & -1 & 0 & 0 & 0 & 0 & 0 & 0 & 0 & 0 & 0 & 0 & 0 \\
0 & 0 & 0 & 0 & 0 & 1 & 0 & 0 & 0 & 0 & 0 & 0 & 0 & 0 & 0 & 0 & 0 & 0 & 0 & 0 & 0 & -1 & 0 & 0 & 0 & 0 & 0 & 0 & 0 & 0 & 0 & 0 \\
0 & 0 & 0 & 0 & 0 & 0 & 1 & 0 & 0 & 0 & 0 & 0 & 0 & 0 & 0 & 0 & 0 & 0 & 0 & 0 & 0 & 0 & -1 & 0 & 0 & 0 & 0 & 0 & 0 & 0 & 0 & 0 \\
0 & 0 & 0 & 0 & 0 & 0 & 0 & 1 & 0 & 0 & 0 & 0 & 0 & 0 & 0 & 0 & 0 & 0 & 0 & 0 & 0 & 0 & 0 & -1 & 0 & 0 & 0 & 0 & 0 & 0 & 0 & 0 \\
0 & 0 & 0 & 0 & 0 & 0 & 0 & 0 & 1 & 0 & 0 & 0 & 0 & 0 & 0 & 0 & 0 & 0 & 0 & 0 & 0 & 0 & 0 & 0 & -1 & 0 & 0 & 0 & 0 & 0 & 0 & 0 \\
0 & 0 & 0 & 0 & 0 & 0 & 0 & 0 & 0 & 1 & 0 & 0 & 0 & 0 & 0 & 0 & 0 & 0 & 0 & 0 & 0 & 0 & 0 & 0 & 0 & -1 & 0 & 0 & 0 & 0 & 0 & 0 \\
0 & 0 & 0 & 0 & 0 & 0 & 0 & 0 & 0 & 0 & 1 & 0 & 0 & 0 & 0 & 0 & 0 & 0 & 0 & 0 & 0 & 0 & 0 & 0 & 0 & 0& -1 & 0 & 0 & 0 & 0 & 0 \\
0 & 0 & 0 & 0 & 0 & 0 & 0 & 0 & 0 & 0 & 0 & 1 & 0 & 0 & 0 & 0 & 0 & 0 & 0 & 0 & 0 & 0 & 0 & 0 & 0 & 0& 0 & -1 & 0 & 0 & 0 & 0 \\
0 & 0 & 0 & 0 & 0 & 0 & 0 & 0 & 0 & 0 & 0 & 0 & 1 & 0 & 0 & 0 & 0 & 0 & 0 & 0 & 0 & 0 & 0 & 0 & 0 & 0& 0 & 0 & -1 & 0 & 0 & 0 \\
0 & 0 & 0 & 0 & 0 & 0 & 0 & 0 & 0 & 0 & 0 & 0 & 0 & 1 & 0 & 0 & 0 & 0 & 0 & 0 & 0 & 0 & 0 & 0 & 0 & 0& 0 & 0 & 0 & -1 & 0 & 0 \\
0 & 0 & 0 & 0 & 0 & 0 & 0 & 0 & 0 & 0 & 0 & 0 & 0 & 0 & 1 & 0 & 0 & 0 & 0 & 0 & 0 & 0 & 0 & 0 & 0 & 0& 0 & 0 & 0 & 0 & -1 & 0 \\
0 & 0 & 0 & 0 & 0 & 0 & 0 & 0 & 0 & 0 & 0 & 0 & 0 & 0 & 0 & 1 & 0 & 0 & 0 & 0 & 0 & 0 & 0 & 0 & 0 & 0& 0 & 0 & 0 & 0 & 0 & -1 \\
\end{bsmallmatrix*}
,
\end{align}

\begin{align}
\mathbf{W}_3
=
\begin{bsmallmatrix*}[r]
\phantom{-}1 & \phantom{-}0 & \phantom{-}0 & \phantom{-}0 & \phantom{-}0 & \phantom{-}0 & \phantom{-}0 & \phantom{-}0 & \phantom{-}1 & \phantom{-}0 & \phantom{-}0 & \phantom{-}0 & \phantom{-}0 & \phantom{-}0 & \phantom{-}0 & \phantom{-}0 & \phantom{-}0 & \phantom{-}0 & \phantom{-}0 & \phantom{-}0 & \phantom{-}0 & \phantom{-}0 & \phantom{-}0 & \phantom{-}0 & \phantom{-}0 & \phantom{-}0 & \phantom{-}0 & \phantom{-}0 & \phantom{-}0 & \phantom{-}0 & \phantom{-}0 & \phantom{-}0 \\
0 & 1 & 0 & 0 & 0 & 0 & 0 & 1 & 0 & 0 & 0 & 0 & 0 & 0 & 0 & 0 & 0 & 0 & 0 & 0 & 0 & 0 & 0 & 0 & 0 & 0& 0 & 0 & 0 & 0 & 0 & 0 \\
0 & 0 & 1 & 0 & 0 & 0 & 1 & 0 & 0 & 0 & 0 & 0 & 0 & 0 & 0 & 0 & 0 & 0 & 0 & 0 & 0 & 0 & 0 & 0 & 0 & 0& 0 & 0 & 0 & 0 & 0 & 0 \\
0 & 0 & 0 & 1 & 0 & 1 & 0 & 0 & 0 & 0 & 0 & 0 & 0 & 0 & 0 & 0 & 0 & 0 & 0 & 0 & 0 & 0 & 0 & 0 & 0 & 0& 0 & 0 & 0 & 0 & 0 & 0 \\
0 & 0 & 0 & 0 & 1 & 0 & 0 & 0 & 0 & 0 & 0 & 0 & 0 & 0 & 0 & 0 & 0 & 0 & 0 & 0 & 0 & 0 & 0 & 0 & 0 & 0& 0 & 0 & 0 & 0 & 0 & 0 \\
0 & 0 & 0 & 1 & 0 & -1 & 0 & 0 & 0 & 0 & 0 & 0 & 0 & 0 & 0 & 0 & 0 & 0 & 0 & 0 & 0 & 0 & 0 & 0 & 0 & 0 & 0 & 0 & 0 & 0 & 0 & 0 \\
0 & 0 & 1 & 0 & 0 & 0 & -1 & 0 & 0 & 0 & 0 & 0 & 0 & 0 & 0 & 0 & 0 & 0 & 0 & 0 & 0 & 0 & 0 & 0 & 0 & 0 & 0 & 0 & 0 & 0 & 0 & 0 \\
0 & 1 & 0 & 0 & 0 & 0 & 0 & -1 & 0 & 0 & 0 & 0 & 0 & 0 & 0 & 0 & 0 & 0 & 0 & 0 & 0 & 0 & 0 & 0 & 0 & 0 & 0 & 0 & 0 & 0 & 0 & 0 \\
1 & 0 & 0 & 0 & 0 & 0 & 0 & 0 & -1 & 0 & 0 & 0 & 0 & 0 & 0 & 0 & 0 & 0 & 0 & 0 & 0 & 0 & 0 & 0 & 0 & 0 & 0 & 0 & 0 & 0 & 0 & 0 \\
0 & 0 & 0 & 0 & 0 & 0 & 0 & 0 & 0 & 1 & 0 & 0 & 0 & 0 & 0 & 1 & 0 & 0 & 0 & 0 & 0 & 0 & 0 & 0 & 0 & 0& 0 & 0 & 0 & 0 & 0 & 0 \\
0 & 0 & 0 & 0 & 0 & 0 & 0 & 0 & 0 & 0 & 1 & 0 & 0 & 0 & 1 & 0 & 0 & 0 & 0 & 0 & 0 & 0 & 0 & 0 & 0 & 0& 0 & 0 & 0 & 0 & 0 & 0 \\
0 & 0 & 0 & 0 & 0 & 0 & 0 & 0 & 0 & 0 & 0 & 1 & 0 & 1 & 0 & 0 & 0 & 0 & 0 & 0 & 0 & 0 & 0 & 0 & 0 & 0& 0 & 0 & 0 & 0 & 0 & 0 \\
0 & 0 & 0 & 0 & 0 & 0 & 0 & 0 & 0 & 0 & 0 & 0 & 1 & 0 & 0 & 0 & 0 & 0 & 0 & 0 & 0 & 0 & 0 & 0 & 0 & 0& 0 & 0 & 0 & 0 & 0 & 0 \\
0 & 0 & 0 & 0 & 0 & 0 & 0 & 0 & 0 & 0 & 0 & 1 & 0 & -1 & 0 & 0 & 0 & 0 & 0 & 0 & 0 & 0 & 0 & 0 & 0 & 0 & 0 & 0 & 0 & 0 & 0 & 0 \\
0 & 0 & 0 & 0 & 0 & 0 & 0 & 0 & 0 & 0 & 1 & 0 & 0 & 0 & -1 & 0 & 0 & 0 & 0 & 0 & 0 & 0 & 0 & 0 & 0 & 0 & 0 & 0 & 0 & 0 & 0 & 0 \\
0 & 0 & 0 & 0 & 0 & 0 & 0 & 0 & 0 & 1 & 0 & 0 & 0 & 0 & 0 & -1 & 0 & 0 & 0 & 0 & 0 & 0 & 0 & 0 & 0 & 0 & 0 & 0 & 0 & 0 & 0 & 0 \\
0 & 0 & 0 & 0 & 0 & 0 & 0 & 0 & 0 & 0 & 0 & 0 & 0 & 0 & 0 & 0 & 1 & 0 & 0 & 0 & 0 & 0 & 0 & 0 & 0 & 0& 0 & 0 & 0 & 0 & 0 & 0 \\
0 & 0 & 0 & 0 & 0 & 0 & 0 & 0 & 0 & 0 & 0 & 0 & 0 & 0 & 0 & 0 & 0 & 1 & 0 & 0 & 0 & 0 & 0 & 0 & 0 & 0& 0 & 0 & 0 & 0 & 0 & 0 \\
0 & 0 & 0 & 0 & 0 & 0 & 0 & 0 & 0 & 0 & 0 & 0 & 0 & 0 & 0 & 0 & 0 & 0 & 1 & 0 & 0 & 0 & 0 & 0 & 0 & 0& 0 & 0 & 0 & 0 & 0 & 0 \\
0 & 0 & 0 & 0 & 0 & 0 & 0 & 0 & 0 & 0 & 0 & 0 & 0 & 0 & 0 & 0 & 0 & 0 & 0 & 1 & 0 & 0 & 0 & 0 & 0 & 0& 0 & 0 & 0 & 0 & 0 & 0 \\
0 & 0 & 0 & 0 & 0 & 0 & 0 & 0 & 0 & 0 & 0 & 0 & 0 & 0 & 0 & 0 & 0 & 0 & 0 & 0 & 1 & 0 & 0 & 0 & 0 & 0& 0 & 0 & 0 & 0 & 0 & 0 \\
0 & 0 & 0 & 0 & 0 & 0 & 0 & 0 & 0 & 0 & 0 & 0 & 0 & 0 & 0 & 0 & 0 & 0 & 0 & 0 & 0 & 1 & 0 & 0 & 0 & 0& 0 & 0 & 0 & 0 & 0 & 0 \\
0 & 0 & 0 & 0 & 0 & 0 & 0 & 0 & 0 & 0 & 0 & 0 & 0 & 0 & 0 & 0 & 0 & 0 & 0 & 0 & 0 & 0 & 1 & 0 & 0 & 0& 0 & 0 & 0 & 0 & 0 & 0 \\
0 & 0 & 0 & 0 & 0 & 0 & 0 & 0 & 0 & 0 & 0 & 0 & 0 & 0 & 0 & 0 & 0 & 0 & 0 & 0 & 0 & 0 & 0 & 1 & 0 & 0& 0 & 0 & 0 & 0 & 0 & 0 \\
0 & 0 & 0 & 0 & 0 & 0 & 0 & 0 & 0 & 0 & 0 & 0 & 0 & 0 & 0 & 0 & 0 & 0 & 0 & 0 & 0 & 0 & 0 & 0 & 1 & 0& 0 & 0 & 0 & 0 & 0 & 0 \\
0 & 0 & 0 & 0 & 0 & 0 & 0 & 0 & 0 & 0 & 0 & 0 & 0 & 0 & 0 & 0 & 0 & 0 & 0 & 0 & 0 & 0 & 0 & 0 & 0 & 1& 0 & 0 & 0 & 0 & 0 & 0 \\
0 & 0 & 0 & 0 & 0 & 0 & 0 & 0 & 0 & 0 & 0 & 0 & 0 & 0 & 0 & 0 & 0 & 0 & 0 & 0 & 0 & 0 & 0 & 0 & 0 & 0& 1 & 0 & 0 & 0 & 0 & 0 \\
0 & 0 & 0 & 0 & 0 & 0 & 0 & 0 & 0 & 0 & 0 & 0 & 0 & 0 & 0 & 0 & 0 & 0 & 0 & 0 & 0 & 0 & 0 & 0 & 0 & 0& 0 & 1 & 0 & 0 & 0 & 0 \\
0 & 0 & 0 & 0 & 0 & 0 & 0 & 0 & 0 & 0 & 0 & 0 & 0 & 0 & 0 & 0 & 0 & 0 & 0 & 0 & 0 & 0 & 0 & 0 & 0 & 0& 0 & 0 & 1 & 0 & 0 & 0 \\
0 & 0 & 0 & 0 & 0 & 0 & 0 & 0 & 0 & 0 & 0 & 0 & 0 & 0 & 0 & 0 & 0 & 0 & 0 & 0 & 0 & 0 & 0 & 0 & 0 & 0& 0 & 0 & 0 & 1 & 0 & 0 \\
0 & 0 & 0 & 0 & 0 & 0 & 0 & 0 & 0 & 0 & 0 & 0 & 0 & 0 & 0 & 0 & 0 & 0 & 0 & 0 & 0 & 0 & 0 & 0 & 0 & 0& 0 & 0 & 0 & 0 & 1 & 0 \\
0 & 0 & 0 & 0 & 0 & 0 & 0 & 0 & 0 & 0 & 0 & 0 & 0 & 0 & 0 & 0 & 0 & 0 & 0 & 0 & 0 & 0 & 0 & 0 & 0 & 0& 0 & 0 & 0 & 0 & 0 & 1 \\
\end{bsmallmatrix*}
,
\end{align}

\begin{align}
\mathbf{W}_4
=
\begin{bsmallmatrix*}[r]
\phantom{-}1 & \phantom{-}0 & \phantom{-}0 & \phantom{-}0 & \phantom{-}1 & \phantom{-}0 & \phantom{-}0 & \phantom{-}0 & \phantom{-}0 & \phantom{-}0 & \phantom{-}0 & \phantom{-}0 & \phantom{-}0 & \phantom{-}0 & \phantom{-}0 & \phantom{-}0 & \phantom{-}0 & \phantom{-}0 & \phantom{-}0 & \phantom{-}0 & \phantom{-}0 & \phantom{-}0 & \phantom{-}0 & \phantom{-}0 & \phantom{-}0 & \phantom{-}0 & \phantom{-}0 & \phantom{-}0 & \phantom{-}0 & \phantom{-}0 & \phantom{-}0 & \phantom{-}0 \\
0 & 1 & 0 & 1 & 0 & 0 & 0 & 0 & 0 & 0 & 0 & 0 & 0 & 0 & 0 & 0 & 0 & 0 & 0 & 0 & 0 & 0 & 0 & 0 & 0 & 0& 0 & 0 & 0 & 0 & 0 & 0 \\
0 & 0 & 1 & 0 & 0 & 0 & 0 & 0 & 0 & 0 & 0 & 0 & 0 & 0 & 0 & 0 & 0 & 0 & 0 & 0 & 0 & 0 & 0 & 0 & 0 & 0& 0 & 0 & 0 & 0 & 0 & 0 \\
0 & 1 & 0 & -1 & 0 & 0 & 0 & 0 & 0 & 0 & 0 & 0 & 0 & 0 & 0 & 0 & 0 & 0 & 0 & 0 & 0 & 0 & 0 & 0 & 0 & 0 & 0 & 0 & 0 & 0 & 0 & 0 \\
1 & 0 & 0 & 0 & -1 & 0 & 0 & 0 & 0 & 0 & 0 & 0 & 0 & 0 & 0 & 0 & 0 & 0 & 0 & 0 & 0 & 0 & 0 & 0 & 0 & 0 & 0 & 0 & 0 & 0 & 0 & 0 \\
0 & 0 & 0 & 0 & 0 & 1 & 0 & 0 & 0 & 0 & 0 & 0 & 0 & 0 & 0 & 0 & 0 & 0 & 0 & 0 & 0 & 0 & 0 & 0 & 0 & 0& 0 & 0 & 0 & 0 & 0 & 0 \\
0 & 0 & 0 & 0 & 0 & 0 & 1 & 0 & 1 & 0 & 0 & 0 & 0 & 0 & 0 & 0 & 0 & 0 & 0 & 0 & 0 & 0 & 0 & 0 & 0 & 0& 0 & 0 & 0 & 0 & 0 & 0 \\
0 & 0 & 0 & 0 & 0 & 0 & 0 & 1 & 0 & 0 & 0 & 0 & 0 & 0 & 0 & 0 & 0 & 0 & 0 & 0 & 0 & 0 & 0 & 0 & 0 & 0& 0 & 0 & 0 & 0 & 0 & 0 \\
0 & 0 & 0 & 0 & 0 & 0 & 1 & 0 & -1 & 0 & 0 & 0 & 0 & 0 & 0 & 0 & 0 & 0 & 0 & 0 & 0 & 0 & 0 & 0 & 0 & 0 & 0 & 0 & 0 & 0 & 0 & 0 \\
0 & 0 & 0 & 0 & 0 & 0 & 0 & 0 & 0 & 1 & 0 & 0 & 0 & 0 & 0 & 0 & 0 & 0 & 0 & 0 & 0 & 0 & 0 & 0 & 0 & 0& 0 & 0 & 0 & 0 & 0 & 0 \\
0 & 0 & 0 & 0 & 0 & 0 & 0 & 0 & 0 & 0 & 1 & 0 & 1 & 0 & 0 & 0 & 0 & 0 & 0 & 0 & 0 & 0 & 0 & 0 & 0 & 0& 0 & 0 & 0 & 0 & 0 & 0 \\
0 & 0 & 0 & 0 & 0 & 0 & 0 & 0 & 0 & 0 & 0 & 1 & 0 & 0 & 0 & 0 & 0 & 0 & 0 & 0 & 0 & 0 & 0 & 0 & 0 & 0& 0 & 0 & 0 & 0 & 0 & 0 \\
0 & 0 & 0 & 0 & 0 & 0 & 0 & 0 & 0 & 0 & 1 & 0 & -1 & 0 & 0 & 0 & 0 & 0 & 0 & 0 & 0 & 0 & 0 & 0 & 0 & 0 & 0 & 0 & 0 & 0 & 0 & 0 \\
0 & 0 & 0 & 0 & 0 & 0 & 0 & 0 & 0 & 0 & 0 & 0 & 0 & 1 & 0 & 1 & 0 & 0 & 0 & 0 & 0 & 0 & 0 & 0 & 0 & 0& 0 & 0 & 0 & 0 & 0 & 0 \\
0 & 0 & 0 & 0 & 0 & 0 & 0 & 0 & 0 & 0 & 0 & 0 & 0 & 0 & 1 & 0 & 0 & 0 & 0 & 0 & 0 & 0 & 0 & 0 & 0 & 0& 0 & 0 & 0 & 0 & 0 & 0 \\
0 & 0 & 0 & 0 & 0 & 0 & 0 & 0 & 0 & 0 & 0 & 0 & 0 & 1 & 0 & -1 & 0 & 0 & 0 & 0 & 0 & 0 & 0 & 0 & 0 & 0 & 0 & 0 & 0 & 0 & 0 & 0 \\
0 & 0 & 0 & 0 & 0 & 0 & 0 & 0 & 0 & 0 & 0 & 0 & 0 & 0 & 0 & 0 & 1 & 0 & 0 & 0 & 0 & 0 & 0 & 0 & 0 & 0& 0 & 0 & 1 & 0 & 0 & 0 \\
0 & 0 & 0 & 0 & 0 & 0 & 0 & 0 & 0 & 0 & 0 & 0 & 0 & 0 & 0 & 0 & 0 & 1 & 0 & 0 & 0 & 0 & 0 & 0 & 0 & 0& 0 & 0 & 0 & 0 & 0 & 0 \\
0 & 0 & 0 & 0 & 0 & 0 & 0 & 0 & 0 & 0 & 0 & 0 & 0 & 0 & 0 & 0 & 0 & 0 & 1 & 0 & 0 & 0 & 0 & 0 & 0 & 0& 0 & 0 & 0 & 0 & 0 & 0 \\
0 & 0 & 0 & 0 & 0 & 0 & 0 & 0 & 0 & 0 & 0 & 0 & 0 & 0 & 0 & 0 & 0 & 0 & 0 & 1 & 0 & 0 & 0 & 0 & 0 & 0& 0 & 0 & 0 & 0 & 0 & 0 \\
0 & 0 & 0 & 0 & 0 & 0 & 0 & 0 & 0 & 0 & 0 & 0 & 0 & 0 & 0 & 0 & 0 & 0 & 0 & 0 & 1 & 0 & 0 & 0 & 1 & 0& 0 & 0 & 0 & 0 & 0 & 0 \\
0 & 0 & 0 & 0 & 0 & 0 & 0 & 0 & 0 & 0 & 0 & 0 & 0 & 0 & 0 & 0 & 0 & 0 & 0 & 0 & 0 & 1 & 0 & 0 & 0 & 0& 0 & 0 & 0 & 0 & 0 & 0 \\
0 & 0 & 0 & 0 & 0 & 0 & 0 & 0 & 0 & 0 & 0 & 0 & 0 & 0 & 0 & 0 & 0 & 0 & 0 & 0 & 0 & 0 & 1 & 0 & 0 & 0& 0 & 0 & 0 & 0 & 0 & 0 \\
0 & 0 & 0 & 0 & 0 & 0 & 0 & 0 & 0 & 0 & 0 & 0 & 0 & 0 & 0 & 0 & 0 & 0 & 0 & 0 & 0 & 0 & 0 & 1 & 0 & 0& 0 & 0 & 0 & 0 & 0 & 0 \\
0 & 0 & 0 & 0 & 0 & 0 & 0 & 0 & 0 & 0 & 0 & 0 & 0 & 0 & 0 & 0 & 0 & 0 & 0 & 0 & 1 & 0 & 0 & 0 & -1 & 0 & 0 & 0 & 0 & 0 & 0 & 0 \\
0 & 0 & 0 & 0 & 0 & 0 & 0 & 0 & 0 & 0 & 0 & 0 & 0 & 0 & 0 & 0 & 0 & 0 & 0 & 0 & 0 & 0 & 0 & 0 & 0 & 1& 0 & 0 & 0 & 0 & 0 & 0 \\
0 & 0 & 0 & 0 & 0 & 0 & 0 & 0 & 0 & 0 & 0 & 0 & 0 & 0 & 0 & 0 & 0 & 0 & 0 & 0 & 0 & 0 & 0 & 0 & 0 & 0& 1 & 0 & 0 & 0 & 0 & 0 \\
0 & 0 & 0 & 0 & 0 & 0 & 0 & 0 & 0 & 0 & 0 & 0 & 0 & 0 & 0 & 0 & 0 & 0 & 0 & 0 & 0 & 0 & 0 & 0 & 0 & 0& 0 & 1 & 0 & 0 & 0 & 0 \\
0 & 0 & 0 & 0 & 0 & 0 & 0 & 0 & 0 & 0 & 0 & 0 & 0 & 0 & 0 & 0 & 1 & 0 & 0 & 0 & 0 & 0 & 0 & 0 & 0 & 0& 0 & 0 & -1 & 0 & 0 & 0 \\
0 & 0 & 0 & 0 & 0 & 0 & 0 & 0 & 0 & 0 & 0 & 0 & 0 & 0 & 0 & 0 & 0 & 0 & 0 & 0 & 0 & 0 & 0 & 0 & 0 & 0& 0 & 0 & 0 & 1 & 0 & 0 \\
0 & 0 & 0 & 0 & 0 & 0 & 0 & 0 & 0 & 0 & 0 & 0 & 0 & 0 & 0 & 0 & 0 & 0 & 0 & 0 & 0 & 0 & 0 & 0 & 0 & 0& 0 & 0 & 0 & 0 & 1 & 0 \\
0 & 0 & 0 & 0 & 0 & 0 & 0 & 0 & 0 & 0 & 0 & 0 & 0 & 0 & 0 & 0 & 0 & 0 & 0 & 0 & 0 & 0 & 0 & 0 & 0 & 0& 0 & 0 & 0 & 0 & 0 & 1 \\
\end{bsmallmatrix*}
,
\end{align}

\begin{align}
\mathbf{W}_5
=
\begin{bsmallmatrix*}[r]
\phantom{-}1 & \phantom{-}0 & \phantom{-}1 & \phantom{-}0 & \phantom{-}0 & \phantom{-}0 & \phantom{-}0 & \phantom{-}0 & \phantom{-}0 & \phantom{-}0 & \phantom{-}0 & \phantom{-}0 & \phantom{-}0 & \phantom{-}0 & \phantom{-}0 & \phantom{-}0 & \phantom{-}0 & \phantom{-}0 & \phantom{-}0 & \phantom{-}0 & \phantom{-}0 & \phantom{-}0 & \phantom{-}0 & \phantom{-}0 & \phantom{-}0 & \phantom{-}0 & \phantom{-}0 & \phantom{-}0 & \phantom{-}0 & \phantom{-}0 & \phantom{-}0 & \phantom{-}0 \\
0 & 1 & 0 & 0 & 0 & 0 & 0 & 0 & 0 & 0 & 0 & 0 & 0 & 0 & 0 & 0 & 0 & 0 & 0 & 0 & 0 & 0 & 0 & 0 & 0 & 0& 0 & 0 & 0 & 0 & 0 & 0 \\
1 & 0 & -1 & 0 & 0 & 0 & 0 & 0 & 0 & 0 & 0 & 0 & 0 & 0 & 0 & 0 & 0 & 0 & 0 & 0 & 0 & 0 & 0 & 0 & 0 & 0 & 0 & 0 & 0 & 0 & 0 & 0 \\
0 & 0 & 0 & 1 & 1 & 0 & 0 & 0 & 0 & 0 & 0 & 0 & 0 & 0 & 0 & 0 & 0 & 0 & 0 & 0 & 0 & 0 & 0 & 0 & 0 & 0& 0 & 0 & 0 & 0 & 0 & 0 \\
0 & 0 & 0 & 1 & -1 & 0 & 0 & 0 & 0 & 0 & 0 & 0 & 0 & 0 & 0 & 0 & 0 & 0 & 0 & 0 & 0 & 0 & 0 & 0 & 0 & 0 & 0 & 0 & 0 & 0 & 0 & 0 \\
0 & 0 & 0 & 0 & 0 & 1 & 0 & 0 & 1 & 0 & 0 & 0 & 0 & 0 & 0 & 0 & 0 & 0 & 0 & 0 & 0 & 0 & 0 & 0 & 0 & 0& 0 & 0 & 0 & 0 & 0 & 0 \\
0 & 0 & 0 & 0 & 0 & 0 & 1 & 1 & 0 & 0 & 0 & 0 & 0 & 0 & 0 & 0 & 0 & 0 & 0 & 0 & 0 & 0 & 0 & 0 & 0 & 0& 0 & 0 & 0 & 0 & 0 & 0 \\
0 & 0 & 0 & 0 & 0 & 0 & 1 & -1 & 0 & 0 & 0 & 0 & 0 & 0 & 0 & 0 & 0 & 0 & 0 & 0 & 0 & 0 & 0 & 0 & 0 & 0 & 0 & 0 & 0 & 0 & 0 & 0 \\
0 & 0 & 0 & 0 & 0 & 1 & 0 & 0 & -1 & 0 & 0 & 0 & 0 & 0 & 0 & 0 & 0 & 0 & 0 & 0 & 0 & 0 & 0 & 0 & 0 & 0 & 0 & 0 & 0 & 0 & 0 & 0 \\
0 & 0 & 0 & 0 & 0 & 0 & 0 & 0 & 0 & 1 & 0 & 0 & 1 & 0 & 0 & 0 & 0 & 0 & 0 & 0 & 0 & 0 & 0 & 0 & 0 & 0& 0 & 0 & 0 & 0 & 0 & 0 \\
0 & 0 & 0 & 0 & 0 & 0 & 0 & 0 & 0 & 0 & 1 & 1 & 0 & 0 & 0 & 0 & 0 & 0 & 0 & 0 & 0 & 0 & 0 & 0 & 0 & 0& 0 & 0 & 0 & 0 & 0 & 0 \\
0 & 0 & 0 & 0 & 0 & 0 & 0 & 0 & 0 & 0 & 1 & -1 & 0 & 0 & 0 & 0 & 0 & 0 & 0 & 0 & 0 & 0 & 0 & 0 & 0 & 0 & 0 & 0 & 0 & 0 & 0 & 0 \\
0 & 0 & 0 & 0 & 0 & 0 & 0 & 0 & 0 & 1 & 0 & 0 & -1 & 0 & 0 & 0 & 0 & 0 & 0 & 0 & 0 & 0 & 0 & 0 & 0 & 0 & 0 & 0 & 0 & 0 & 0 & 0 \\
0 & 0 & 0 & 0 & 0 & 0 & 0 & 0 & 0 & 0 & 0 & 0 & 0 & 1 & 1 & 0 & 0 & 0 & 0 & 0 & 0 & 0 & 0 & 0 & 0 & 0& 0 & 0 & 0 & 0 & 0 & 0 \\
0 & 0 & 0 & 0 & 0 & 0 & 0 & 0 & 0 & 0 & 0 & 0 & 0 & 1 & -1 & 0 & 0 & 0 & 0 & 0 & 0 & 0 & 0 & 0 & 0 & 0 & 0 & 0 & 0 & 0 & 0 & 0 \\
0 & 0 & 0 & 0 & 0 & 0 & 0 & 0 & 0 & 0 & 0 & 0 & 0 & 0 & 0 & 1 & 0 & 0 & 0 & 0 & 0 & 0 & 0 & 0 & 0 & 0& 0 & 0 & 0 & 0 & 0 & 0 \\
0 & 0 & 0 & 0 & 0 & 0 & 0 & 0 & 0 & 0 & 0 & 0 & 0 & 0 & 0 & 0 & -1 & 0 & 0 & 0 & 0 & 0 & 0 & 0 & 0 & 0 & 0 & 0 & 0 & 0 & 1 & 0 \\
0 & 0 & 0 & 0 & 0 & 0 & 0 & 0 & 0 & 0 & 0 & 0 & 0 & 0 & 0 & 0 & 0 & 1 & 0 & 0 & 0 & 0 & 0 & 0 & 0 & 0& 0 & 0 & 0 & 0 & 0 & 0 \\
0 & 0 & 0 & 0 & 0 & 0 & 0 & 0 & 0 & 0 & 0 & 0 & 0 & 0 & 0 & 0 & 0 & 0 & 1 & 0 & 0 & 0 & 0 & 0 & 1 & 0& 0 & 0 & 0 & 0 & 0 & 0 \\
0 & 0 & 0 & 0 & 0 & 0 & 0 & 0 & 0 & 0 & 0 & 0 & 0 & 0 & 0 & 0 & 0 & 0 & 0 & 1 & 0 & 1 & 0 & 1 & 0 & 0& 0 & 0 & 0 & 0 & 0 & 0 \\
0 & 0 & 0 & 0 & 0 & 0 & 0 & 0 & 0 & 0 & 0 & 0 & 0 & 0 & 0 & 0 & 0 & 0 & 0 & 0 & 1 & 0 & 1 & 0 & 0 & 0& 0 & 0 & 0 & 0 & 0 & 0 \\
0 & 0 & 0 & 0 & 0 & 0 & 0 & 0 & 0 & 0 & 0 & 0 & 0 & 0 & 0 & 0 & 0 & 0 & 0 & 1 & 0 & -1 & 0 & 0 & 0 & 0 & 0 & 0 & 0 & 0 & 0 & 0 \\
0 & 0 & 0 & 0 & 0 & 0 & 0 & 0 & 0 & 0 & 0 & 0 & 0 & 0 & 0 & 0 & 0 & 0 & 0 & 0 & 1 & 0 & -1 & 0 & 0 & 0 & 0 & 0 & 0 & 0 & 0 & 0 \\
0 & 0 & 0 & 0 & 0 & 0 & 0 & 0 & 0 & 0 & 0 & 0 & 0 & 0 & 0 & 0 & 0 & 0 & 0 & 1 & 0 & 0 & 0 & -1 & 0 & 0 & 0 & 0 & 0 & 0 & 0 & 0 \\
0 & 0 & 0 & 0 & 0 & 0 & 0 & 0 & 0 & 0 & 0 & 0 & 0 & 0 & 0 & 0 & 0 & 0 & 1 & 0 & 0 & 0 & 0 & 0 & -1 & 0 & 0 & 0 & 0 & 0 & 0 & 0 \\
0 & 0 & 0 & 0 & 0 & 0 & 0 & 0 & 0 & 0 & 0 & 0 & 0 & 0 & 0 & 0 & 0 & 0 & 0 & 0 & 0 & 0 & 0 & 0 & 0 & 1& 0 & 0 & 0 & 0 & 0 & 0 \\
0 & 0 & 0 & 0 & 0 & 0 & 0 & 0 & 0 & 0 & 0 & 0 & 0 & 0 & 0 & 0 & 0 & 0 & 0 & 0 & 0 & 0 & 0 & 0 & 0 & 0& 1 & 0 & 1 & 0 & 0 & 0 \\
0 & 0 & 0 & 0 & 0 & 0 & 0 & 0 & 0 & 0 & 0 & 0 & 0 & 0 & 0 & 0 & 0 & 0 & 0 & 0 & 0 & 0 & 0 & 0 & 0 & 0& 0 & 1 & 0 & 1 & 0 & 1 \\
0 & 0 & 0 & 0 & 0 & 0 & 0 & 0 & 0 & 0 & 0 & 0 & 0 & 0 & 0 & 0 & 0 & 0 & 0 & 0 & 0 & 0 & 0 & 0 & 0 & 0& 1 & 0 & -1 & 0 & 0 & 0 \\
0 & 0 & 0 & 0 & 0 & 0 & 0 & 0 & 0 & 0 & 0 & 0 & 0 & 0 & 0 & 0 & 0 & 0 & 0 & 0 & 0 & 0 & 0 & 0 & 0 & 0& 0 & 1 & 0 & -1 & 0 & 0 \\
0 & 0 & 0 & 0 & 0 & 0 & 0 & 0 & 0 & 0 & 0 & 0 & 0 & 0 & 0 & 0 & 1 & 0 & 0 & 0 & 0 & 0 & 0 & 0 & 0 & 0& 0 & 0 & 0 & 0 & 1 & 0 \\
0 & 0 & 0 & 0 & 0 & 0 & 0 & 0 & 0 & 0 & 0 & 0 & 0 & 0 & 0 & 0 & 0 & 0 & 0 & 0 & 0 & 0 & 0 & 0 & 0 & 0& 0 & 1 & 0 & 0 & 0 & -1 \\
\end{bsmallmatrix*}
,
\end{align}

\begin{align}
\mathbf{W}_6
=
\begin{bsmallmatrix*}[r]
\phantom{-}1 & \phantom{-}1 & \phantom{-}0 & \phantom{-}0 & \phantom{-}0 & \phantom{-}0 & \phantom{-}0 & \phantom{-}0 & \phantom{-}0 & \phantom{-}0 & \phantom{-}0 & \phantom{-}0 & \phantom{-}0 & \phantom{-}0 & \phantom{-}0 & \phantom{-}0 & \phantom{-}0 & \phantom{-}0 & \phantom{-}0 & \phantom{-}0 & \phantom{-}0 & \phantom{-}0 & \phantom{-}0 & \phantom{-}0 & \phantom{-}0 & \phantom{-}0 & \phantom{-}0 & \phantom{-}0 & \phantom{-}0 & \phantom{-}0 & \phantom{-}0 & \phantom{-}0 \\
1 & -1 & 0 & 0 & 0 & 0 & 0 & 0 & 0 & 0 & 0 & 0 & 0 & 0 & 0 & 0 & 0 & 0 & 0 & 0 & 0 & 0 & 0 & 0 & 0 & 0 & 0 & 0 & 0 & 0 & 0 & 0 \\
0 & 0 & 1 & 0 & 0 & 0 & 0 & 0 & 0 & 0 & 0 & 0 & 0 & 0 & 0 & 0 & 0 & 0 & 0 & 0 & 0 & 0 & 0 & 0 & 0 & 0& 0 & 0 & 0 & 0 & 0 & 0 \\
0 & 0 & 0 & 1 & 0 & 0 & 0 & 0 & 0 & 0 & 0 & 0 & 0 & 0 & 0 & 0 & 0 & 0 & 0 & 0 & 0 & 0 & 0 & 0 & 0 & 0& 0 & 0 & 0 & 0 & 0 & 0 \\
0 & 0 & 0 & 0 & 1 & 0 & 0 & 0 & 0 & 0 & 0 & 0 & 0 & 0 & 0 & 0 & 0 & 0 & 0 & 0 & 0 & 0 & 0 & 0 & 0 & 0& 0 & 0 & 0 & 0 & 0 & 0 \\
0 & 0 & 0 & 0 & 0 & 1 & 0 & 0 & 0 & 0 & 0 & 0 & 0 & 0 & 0 & 0 & 0 & 0 & 0 & 0 & 0 & 0 & 0 & 0 & 0 & 0& 0 & 0 & 0 & 0 & 0 & 0 \\
0 & 0 & 0 & 0 & 0 & 0 & 1 & 0 & 0 & 0 & 0 & 0 & 0 & 0 & 0 & 0 & 0 & 0 & 0 & 0 & 0 & 0 & 0 & 0 & 0 & 0& 0 & 0 & 0 & 0 & 0 & 0 \\
0 & 0 & 0 & 0 & 0 & 0 & 0 & 1 & 0 & 0 & 0 & 0 & 0 & 0 & 0 & 0 & 0 & 0 & 0 & 0 & 0 & 0 & 0 & 0 & 0 & 0& 0 & 0 & 0 & 0 & 0 & 0 \\
0 & 0 & 0 & 0 & 0 & 0 & 0 & 0 & 1 & 0 & 0 & 0 & 0 & 0 & 0 & 0 & 0 & 0 & 0 & 0 & 0 & 0 & 0 & 0 & 0 & 0& 0 & 0 & 0 & 0 & 0 & 0 \\
0 & 0 & 0 & 0 & 0 & 0 & 0 & 0 & 0 & 1 & 0 & 0 & 0 & 0 & 0 & 0 & 0 & 0 & 0 & 0 & 0 & 0 & 0 & 0 & 0 & 0& 0 & 0 & 0 & 0 & 0 & 0 \\
0 & 0 & 0 & 0 & 0 & 0 & 0 & 0 & 0 & 0 & 1 & 0 & 0 & 0 & 0 & 0 & 0 & 0 & 0 & 0 & 0 & 0 & 0 & 0 & 0 & 0& 0 & 0 & 0 & 0 & 0 & 0 \\
0 & 0 & 0 & 0 & 0 & 0 & 0 & 0 & 0 & 0 & 0 & 1 & 0 & 0 & 0 & 0 & 0 & 0 & 0 & 0 & 0 & 0 & 0 & 0 & 0 & 0& 0 & 0 & 0 & 0 & 0 & 0 \\
0 & 0 & 0 & 0 & 0 & 0 & 0 & 0 & 0 & 0 & 0 & 0 & 1 & 0 & 0 & 0 & 0 & 0 & 0 & 0 & 0 & 0 & 0 & 0 & 0 & 0& 0 & 0 & 0 & 0 & 0 & 0 \\
0 & 0 & 0 & 0 & 0 & 0 & 0 & 0 & 0 & 0 & 0 & 0 & 0 & 1 & 0 & 0 & 0 & 0 & 0 & 0 & 0 & 0 & 0 & 0 & 0 & 0& 0 & 0 & 0 & 0 & 0 & 0 \\
0 & 0 & 0 & 0 & 0 & 0 & 0 & 0 & 0 & 0 & 0 & 0 & 0 & 0 & 1 & 0 & 0 & 0 & 0 & 0 & 0 & 0 & 0 & 0 & 0 & 0& 0 & 0 & 0 & 0 & 0 & 0 \\
0 & 0 & 0 & 0 & 0 & 0 & 0 & 0 & 0 & 0 & 0 & 0 & 0 & 0 & 0 & 1 & 0 & 0 & 0 & 0 & 0 & 0 & 0 & 0 & 0 & 0& 0 & 0 & 0 & 0 & 0 & 0 \\
0 & 0 & 0 & 0 & 0 & 0 & 0 & 0 & 0 & 0 & 0 & 0 & 0 & 0 & 0 & 0 & 1 & 0 & 0 & 0 & 0 & 0 & 0 & 0 & 0 & 0& 0 & 0 & 0 & 0 & 0 & 0 \\
0 & 0 & 0 & 0 & 0 & 0 & 0 & 0 & 0 & 0 & 0 & 0 & 0 & 0 & 0 & 0 & 0 & 1 & 0 & 0 & 0 & 1 & 0 & -1 & 0 & 0 & 0 & 0 & 0 & 0 & 0 & 0 \\
0 & 0 & 0 & 0 & 0 & 0 & 0 & 0 & 0 & 0 & 0 & 0 & 0 & 0 & 0 & 0 & 0 & 0 & 1 & 0 & 0 & 0 & 0 & 0 & 0 & 0& 0 & 0 & 0 & 0 & 0 & 0 \\
0 & 0 & 0 & 0 & 0 & 0 & 0 & 0 & 0 & 0 & 0 & 0 & 0 & 0 & 0 & 0 & 0 & 0 & 0 & 1 & 1 & 0 & 0 & 0 & 0 & 0& 0 & 0 & 0 & 0 & 0 & 0 \\
0 & 0 & 0 & 0 & 0 & 0 & 0 & 0 & 0 & 0 & 0 & 0 & 0 & 0 & 0 & 0 & 0 & 0 & 0 & 1 & -1 & 0 & 0 & 0 & 0 & 0 & 0 & 0 & 0 & 0 & 0 & 0 \\
0 & 0 & 0 & 0 & 0 & 0 & 0 & 0 & 0 & 0 & 0 & 0 & 0 & 0 & 0 & 0 & 0 & 1 & 0 & 0 & 0 & -1 & 0 & 0 & 0 & 0 & 0 & 0 & 0 & 0 & 0 & 0 \\
0 & 0 & 0 & 0 & 0 & 0 & 0 & 0 & 0 & 0 & 0 & 0 & 0 & 0 & 0 & 0 & 0 & 0 & 0 & 0 & 0 & 0 & 1 & 0 & 0 & 0& 0 & 0 & 0 & 0 & 0 & 0 \\
0 & 0 & 0 & 0 & 0 & 0 & 0 & 0 & 0 & 0 & 0 & 0 & 0 & 0 & 0 & 0 & 0 & 1 & 0 & 0 & 0 & 0 & 0 & 1 & 0 & 0& 0 & 0 & 0 & 0 & 0 & 0 \\
0 & 0 & 0 & 0 & 0 & 0 & 0 & 0 & 0 & 0 & 0 & 0 & 0 & 0 & 0 & 0 & 0 & 0 & 0 & 0 & 0 & 0 & 0 & 0 & 1 & 0& 0 & 0 & 0 & 0 & 0 & 0 \\
0 & 0 & 0 & 0 & 0 & 0 & 0 & 0 & 0 & 0 & 0 & 0 & 0 & 0 & 0 & 0 & 0 & 0 & 0 & 0 & 0 & 0 & 0 & 0 & 0 & 1& 0 & 0 & 0 & 1 & 0 & -1 \\
0 & 0 & 0 & 0 & 0 & 0 & 0 & 0 & 0 & 0 & 0 & 0 & 0 & 0 & 0 & 0 & 0 & 0 & 0 & 0 & 0 & 0 & 0 & 0 & 0 & 0& 1 & 0 & 0 & 0 & 0 & 0 \\
0 & 0 & 0 & 0 & 0 & 0 & 0 & 0 & 0 & 0 & 0 & 0 & 0 & 0 & 0 & 0 & 0 & 0 & 0 & 0 & 0 & 0 & 0 & 0 & 0 & 0& 0 & 1 & 0 & 0 & 1 & 0 \\
0 & 0 & 0 & 0 & 0 & 0 & 0 & 0 & 0 & 0 & 0 & 0 & 0 & 0 & 0 & 0 & 0 & 0 & 0 & 0 & 0 & 0 & 0 & 0 & 0 & 0& 0 & 0 & 1 & 0 & 0 & 0 \\
0 & 0 & 0 & 0 & 0 & 0 & 0 & 0 & 0 & 0 & 0 & 0 & 0 & 0 & 0 & 0 & 0 & 0 & 0 & 0 & 0 & 0 & 0 & 0 & 0 & 1& 0 & 0 & 0 & -1 & 0 & 0 \\
0 & 0 & 0 & 0 & 0 & 0 & 0 & 0 & 0 & 0 & 0 & 0 & 0 & 0 & 0 & 0 & 0 & 0 & 0 & 0 & 0 & 0 & 0 & 0 & 0 & 0& 0 & 1 & 0 & 0 & -1 & 0 \\
0 & 0 & 0 & 0 & 0 & 0 & 0 & 0 & 0 & 0 & 0 & 0 & 0 & 0 & 0 & 0 & 0 & 0 & 0 & 0 & 0 & 0 & 0 & 0 & 0 & 1& 0 & 0 & 0 & 0 & 0 & 1 \\
\end{bsmallmatrix*}
,
\end{align}

\begin{align}
\mathbf{W}_7
=
\begin{bsmallmatrix*}[r]
\phantom{-}1 & \phantom{-}0 & \phantom{-}0 & \phantom{-}0 & \phantom{-}0 & \phantom{-}0 & \phantom{-}0 & \phantom{-}0 & \phantom{-}0 & \phantom{-}0 & \phantom{-}0 & \phantom{-}0 & \phantom{-}0 & \phantom{-}0 & \phantom{-}0 & \phantom{-}0 & \phantom{-}0 & \phantom{-}0 & \phantom{-}0 & \phantom{-}0 & \phantom{-}0 & \phantom{-}0 & \phantom{-}0 & \phantom{-}0 & \phantom{-}0 & \phantom{-}0 & \phantom{-}0 & \phantom{-}0 & \phantom{-}0 & \phantom{-}0 & \phantom{-}0 & \phantom{-}0 \\
0 & 1 & 0 & 0 & 0 & 0 & 0 & 0 & 0 & 0 & 0 & 0 & 0 & 0 & 0 & 0 & 0 & 0 & 0 & 0 & 0 & 0 & 0 & 0 & 0 & 0& 0 & 0 & 0 & 0 & 0 & 0 \\
0 & 0 & 1 & 0 & 0 & 0 & 0 & 0 & 0 & 0 & 0 & 0 & 0 & 0 & 0 & 0 & 0 & 0 & 0 & 0 & 0 & 0 & 0 & 0 & 0 & 0& 0 & 0 & 0 & 0 & 0 & 0 \\
0 & 0 & 0 & 1 & 0 & 0 & 0 & 0 & 0 & 0 & 0 & 0 & 0 & 0 & 0 & 0 & 0 & 0 & 0 & 0 & 0 & 0 & 0 & 0 & 0 & 0& 0 & 0 & 0 & 0 & 0 & 0 \\
0 & 0 & 0 & 0 & 1 & 0 & 0 & 0 & 0 & 0 & 0 & 0 & 0 & 0 & 0 & 0 & 0 & 0 & 0 & 0 & 0 & 0 & 0 & 0 & 0 & 0& 0 & 0 & 0 & 0 & 0 & 0 \\
0 & 0 & 0 & 0 & 0 & 1 & 0 & 0 & 0 & 0 & 0 & 0 & 0 & 0 & 0 & 0 & 0 & 0 & 0 & 0 & 0 & 0 & 0 & 0 & 0 & 0& 0 & 0 & 0 & 0 & 0 & 0 \\
0 & 0 & 0 & 0 & 0 & 0 & 1 & 0 & 0 & 0 & 0 & 0 & 0 & 0 & 0 & 0 & 0 & 0 & 0 & 0 & 0 & 0 & 0 & 0 & 0 & 0& 0 & 0 & 0 & 0 & 0 & 0 \\
0 & 0 & 0 & 0 & 0 & 0 & 0 & 1 & 0 & 0 & 0 & 0 & 0 & 0 & 0 & 0 & 0 & 0 & 0 & 0 & 0 & 0 & 0 & 0 & 0 & 0& 0 & 0 & 0 & 0 & 0 & 0 \\
0 & 0 & 0 & 0 & 0 & 0 & 0 & 0 & 1 & 0 & 0 & 0 & 0 & 0 & 0 & 0 & 0 & 0 & 0 & 0 & 0 & 0 & 0 & 0 & 0 & 0& 0 & 0 & 0 & 0 & 0 & 0 \\
0 & 0 & 0 & 0 & 0 & 0 & 0 & 0 & 0 & 1 & 0 & 0 & 0 & 0 & 0 & 0 & 0 & 0 & 0 & 0 & 0 & 0 & 0 & 0 & 0 & 0& 0 & 0 & 0 & 0 & 0 & 0 \\
0 & 0 & 0 & 0 & 0 & 0 & 0 & 0 & 0 & 0 & 1 & 0 & 0 & 0 & 0 & 0 & 0 & 0 & 0 & 0 & 0 & 0 & 0 & 0 & 0 & 0& 0 & 0 & 0 & 0 & 0 & 0 \\
0 & 0 & 0 & 0 & 0 & 0 & 0 & 0 & 0 & 0 & 0 & 1 & 0 & 0 & 0 & 0 & 0 & 0 & 0 & 0 & 0 & 0 & 0 & 0 & 0 & 0& 0 & 0 & 0 & 0 & 0 & 0 \\
0 & 0 & 0 & 0 & 0 & 0 & 0 & 0 & 0 & 0 & 0 & 0 & 1 & 0 & 0 & 0 & 0 & 0 & 0 & 0 & 0 & 0 & 0 & 0 & 0 & 0& 0 & 0 & 0 & 0 & 0 & 0 \\
0 & 0 & 0 & 0 & 0 & 0 & 0 & 0 & 0 & 0 & 0 & 0 & 0 & 1 & 0 & 0 & 0 & 0 & 0 & 0 & 0 & 0 & 0 & 0 & 0 & 0& 0 & 0 & 0 & 0 & 0 & 0 \\
0 & 0 & 0 & 0 & 0 & 0 & 0 & 0 & 0 & 0 & 0 & 0 & 0 & 0 & 1 & 0 & 0 & 0 & 0 & 0 & 0 & 0 & 0 & 0 & 0 & 0& 0 & 0 & 0 & 0 & 0 & 0 \\
0 & 0 & 0 & 0 & 0 & 0 & 0 & 0 & 0 & 0 & 0 & 0 & 0 & 0 & 0 & 1 & 0 & 0 & 0 & 0 & 0 & 0 & 0 & 0 & 0 & 0& 0 & 0 & 0 & 0 & 0 & 0 \\
0 & 0 & 0 & 0 & 0 & 0 & 0 & 0 & 0 & 0 & 0 & 0 & 0 & 0 & 0 & 0 & 1 & 0 & 0 & 0 & 0 & 0 & 0 & 0 & 0 & 0& 0 & 0 & 0 & 1 & 0 & 0 \\
0 & 0 & 0 & 0 & 0 & 0 & 0 & 0 & 0 & 0 & 0 & 0 & 0 & 0 & 0 & 0 & 0 & 1 & 0 & 0 & 0 & 0 & 0 & 0 & 1 & 0& 0 & 0 & 0 & 0 & 0 & 0 \\
0 & 0 & 0 & 0 & 0 & 0 & 0 & 0 & 0 & 0 & 0 & 0 & 0 & 0 & 0 & 0 & 0 & 0 & -1 & 0 & 0 & 0 & 0 & 1 & 0 & 0 & 0 & 0 & 0 & 0 & 0 & 0 \\
0 & 0 & 0 & 0 & 0 & 0 & 0 & 0 & 0 & 0 & 0 & 0 & 0 & 0 & 0 & 0 & 0 & 0 & 0 & 1 & 0 & 0 & 0 & 0 & 0 & 0& 0 & 0 & 0 & 0 & 0 & 0 \\
0 & 0 & 0 & 0 & 0 & 0 & 0 & 0 & 0 & 0 & 0 & 0 & 0 & 0 & 0 & 0 & 0 & 0 & 0 & 0 & 1 & 0 & 0 & 0 & 0 & 0& 0 & 0 & 0 & 0 & 0 & 0 \\
0 & 0 & 0 & 0 & 0 & 0 & 0 & 0 & 0 & 0 & 0 & 0 & 0 & 0 & 0 & 0 & 0 & 0 & 0 & 0 & 0 & 1 & 1 & 0 & 0 & 0& 0 & 0 & 0 & 0 & 0 & 0 \\
0 & 0 & 0 & 0 & 0 & 0 & 0 & 0 & 0 & 0 & 0 & 0 & 0 & 0 & 0 & 0 & 0 & 0 & 0 & 0 & 0 & 1 & -1 & 0 & 0 & 0 & 0 & 0 & 0 & 0 & 0 & 0 \\
0 & 0 & 0 & 0 & 0 & 0 & 0 & 0 & 0 & 0 & 0 & 0 & 0 & 0 & 0 & 0 & 0 & 0 & 1 & 0 & 0 & 0 & 0 & 1 & 0 & 0& 0 & 0 & 0 & 0 & 0 & 0 \\
0 & 0 & 0 & 0 & 0 & 0 & 0 & 0 & 0 & 0 & 0 & 0 & 0 & 0 & 0 & 0 & 0 & 1 & 0 & 0 & 0 & 0 & 0 & 0 & -1 & 0 & 0 & 0 & 0 & 0 & 0 & 0 \\
0 & 0 & 0 & 0 & 0 & 0 & 0 & 0 & 0 & 0 & 0 & 0 & 0 & 0 & 0 & 0 & 0 & 0 & 0 & 0 & 0 & 0 & 0 & 0 & 0 & 1& 1 & 0 & 0 & 0 & 0 & 0 \\
0 & 0 & 0 & 0 & 0 & 0 & 0 & 0 & 0 & 0 & 0 & 0 & 0 & 0 & 0 & 0 & 0 & 0 & 0 & 0 & 0 & 0 & 0 & 0 & 0 & 1& -1 & 0 & 0 & 0 & 0 & 0 \\
0 & 0 & 0 & 0 & 0 & 0 & 0 & 0 & 0 & 0 & 0 & 0 & 0 & 0 & 0 & 0 & 0 & 0 & 0 & 0 & 0 & 0 & 0 & 0 & 0 & 0& 0 & 1 & 0 & 0 & 0 & 0 \\
0 & 0 & 0 & 0 & 0 & 0 & 0 & 0 & 0 & 0 & 0 & 0 & 0 & 0 & 0 & 0 & 0 & 0 & 0 & 0 & 0 & 0 & 0 & 0 & 0 & 0& 0 & 0 & 1 & 0 & 0 & 1 \\
0 & 0 & 0 & 0 & 0 & 0 & 0 & 0 & 0 & 0 & 0 & 0 & 0 & 0 & 0 & 0 & 1 & 0 & 0 & 0 & 0 & 0 & 0 & 0 & 0 & 0& 0 & 0 & 0 & -1 & 0 & 0 \\
0 & 0 & 0 & 0 & 0 & 0 & 0 & 0 & 0 & 0 & 0 & 0 & 0 & 0 & 0 & 0 & 0 & 0 & 0 & 0 & 0 & 0 & 0 & 0 & 0 & 0& 0 & 0 & 0 & 0 & 1 & 0 \\
0 & 0 & 0 & 0 & 0 & 0 & 0 & 0 & 0 & 0 & 0 & 0 & 0 & 0 & 0 & 0 & 0 & 0 & 0 & 0 & 0 & 0 & 0 & 0 & 0 & 0& 0 & 0 & 1 & 0 & 0 & -1 \\
\end{bsmallmatrix*}
,
\end{align}

\begin{align}
\mathbf{W}_8
=
\begin{bsmallmatrix*}[r]
\phantom{-}1 & \phantom{-}0 & \phantom{-}0 & \phantom{-}0 & \phantom{-}0 & \phantom{-}0 & \phantom{-}0 & \phantom{-}0 & \phantom{-}0 & \phantom{-}0 & \phantom{-}0 & \phantom{-}0 & \phantom{-}0 & \phantom{-}0 & \phantom{-}0 & \phantom{-}0 & \phantom{-}0 & \phantom{-}0 & \phantom{-}0 & \phantom{-}0 & \phantom{-}0 & \phantom{-}0 & \phantom{-}0 & \phantom{-}0 & \phantom{-}0 & \phantom{-}0 & \phantom{-}0 & \phantom{-}0 & \phantom{-}0 & \phantom{-}0 & \phantom{-}0 & \phantom{-}0 \\
0 & 0 & 0 & 0 & 0 & 0 & 0 & 0 & 0 & 0 & 0 & 0 & 0 & 0 & 0 & 0 & 0 & 0 & 0 & -j & 0 & 0 & 0 & 0 & 0& 0 & 0 & 1 & 0 & 0 & 0 & 0 \\
0 & 0 & 0 & 0 & 0 & 0 & 1 & 0 & 0 & 0 & -j & 0 & 0 & 0 & 0 & 0 & 0 & 0 & 0 & 0 & 0 & 0 & 0 & 0 & 0& 0 & 0 & 0 & 0 & 0 & 0 & 0 \\
0 & 0 & 0 & 0 & 0 & 0 & 0 & 0 & 0 & 0 & 0 & 0 & 0 & 0 & 0 & 0 & 0 & 0 & 0 & 0 & 0 & 0 & 0 & -j & 0& 0 & 0 & 0 & -1 & 0 & 0 & 0 \\
0 & 0 & 0 & 1 & 0 & 0 & 0 & 0 & 0 & 0 & 0 & 0 & 0 & j & 0 & 0 & 0 & 0 & 0 & 0 & 0 & 0 & 0 & 0 & 0 & 0 & 0 & 0 & 0 & 0 & 0 & 0 \\
0 & 0 & 0 & 0 & 0 & 0 & 0 & 0 & 0 & 0 & 0 & 0 & 0 & 0 & 0 & 0 & 0 & -j & 0 & 0 & 0 & 0 & 0 & 0 & 0& 1 & 0 & 0 & 0 & 0 & 0 & 0 \\
0 & 0 & 0 & 0 & 0 & -1 & 0 & 0 & 0 & -j & 0 & 0 & 0 & 0 & 0 & 0 & 0 & 0 & 0 & 0 & 0 & 0 & 0 & 0 & 0 & 0 & 0 & 0 & 0 & 0 & 0 & 0 \\
0 & 0 & 0 & 0 & 0 & 0 & 0 & 0 & 0 & 0 & 0 & 0 & 0 & 0 & 0 & 0 & -1 & 0 & 0 & 0 & 0 & 0 & -j & 0 & 0 & 0 & 0 & 0 & 0 & 0 & 0 & 0 \\
0 & 0 & 1 & 0 & 0 & 0 & 0 & 0 & 0 & 0 & 0 & 0 & 0 & 0 & 0 & -j & 0 & 0 & 0 & 0 & 0 & 0 & 0 & 0 & 0& 0 & 0 & 0 & 0 & 0 & 0 & 0 \\
0 & 0 & 0 & 0 & 0 & 0 & 0 & 0 & 0 & 0 & 0 & 0 & 0 & 0 & 0 & 0 & 0 & 0 & 0 & 0 & 0 & -j & 0 & 0 & 0& 0 & 0 & 0 & 0 & -1 & 0 & 0 \\
0 & 0 & 0 & 0 & 0 & 0 & 0 & 0 & 1 & 0 & 0 & 0 & -j & 0 & 0 & 0 & 0 & 0 & 0 & 0 & 0 & 0 & 0 & 0 & 0& 0 & 0 & 0 & 0 & 0 & 0 & 0 \\
0 & 0 & 0 & 0 & 0 & 0 & 0 & 0 & 0 & 0 & 0 & 0 & 0 & 0 & 0 & 0 & 0 & 0 & 0 & 0 & 0 & 0 & 0 & 0 & -j& 0 & -1 & 0 & 0 & 0 & 0 & 0 \\
0 & 0 & 0 & 0 & -1 & 0 & 0 & 0 & 0 & 0 & 0 & 0 & 0 & 0 & j & 0 & 0 & 0 & 0 & 0 & 0 & 0 & 0 & 0 & 0& 0 & 0 & 0 & 0 & 0 & 0 & 0 \\
0 & 0 & 0 & 0 & 0 & 0 & 0 & 0 & 0 & 0 & 0 & 0 & 0 & 0 & 0 & 0 & 0 & 0 & -j & 0 & 0 & 0 & 0 & 0 & 0& 0 & 0 & 0 & 0 & 0 & 0 & -1 \\
0 & 0 & 0 & 0 & 0 & 0 & 0 & 1 & 0 & 0 & 0 & j & 0 & 0 & 0 & 0 & 0 & 0 & 0 & 0 & 0 & 0 & 0 & 0 & 0 & 0 & 0 & 0 & 0 & 0 & 0 & 0 \\
0 & 0 & 0 & 0 & 0 & 0 & 0 & 0 & 0 & 0 & 0 & 0 & 0 & 0 & 0 & 0 & 0 & 0 & 0 & 0 & -j & 0 & 0 & 0 & 0& 0 & 0 & 0 & 0 & 0 & -1 & 0 \\
0 & 1 & 0 & 0 & 0 & 0 & 0 & 0 & 0 & 0 & 0 & 0 & 0 & 0 & 0 & 0 & 0 & 0 & 0 & 0 & 0 & 0 & 0 & 0 & 0 & 0& 0 & 0 & 0 & 0 & 0 & 0 \\
0 & 0 & 0 & 0 & 0 & 0 & 0 & 0 & 0 & 0 & 0 & 0 & 0 & 0 & 0 & 0 & 0 & 0 & 0 & 0 & j & 0 & 0 & 0 & 0 & 0 & 0 & 0 & 0 & 0 & -1 & 0 \\
0 & 0 & 0 & 0 & 0 & 0 & 0 & 1 & 0 & 0 & 0 & -j & 0 & 0 & 0 & 0 & 0 & 0 & 0 & 0 & 0 & 0 & 0 & 0 & 0& 0 & 0 & 0 & 0 & 0 & 0 & 0 \\
0 & 0 & 0 & 0 & 0 & 0 & 0 & 0 & 0 & 0 & 0 & 0 & 0 & 0 & 0 & 0 & 0 & 0 & j & 0 & 0 & 0 & 0 & 0 & 0 & 0 & 0 & 0 & 0 & 0 & 0 & -1 \\
0 & 0 & 0 & 0 & -1 & 0 & 0 & 0 & 0 & 0 & 0 & 0 & 0 & 0 & -j & 0 & 0 & 0 & 0 & 0 & 0 & 0 & 0 & 0 & 0 & 0 & 0 & 0 & 0 & 0 & 0 & 0 \\
0 & 0 & 0 & 0 & 0 & 0 & 0 & 0 & 0 & 0 & 0 & 0 & 0 & 0 & 0 & 0 & 0 & 0 & 0 & 0 & 0 & 0 & 0 & 0 & j & 0 & -1 & 0 & 0 & 0 & 0 & 0 \\
0 & 0 & 0 & 0 & 0 & 0 & 0 & 0 & 1 & 0 & 0 & 0 & j & 0 & 0 & 0 & 0 & 0 & 0 & 0 & 0 & 0 & 0 & 0 & 0 & 0 & 0 & 0 & 0 & 0 & 0 & 0 \\
0 & 0 & 0 & 0 & 0 & 0 & 0 & 0 & 0 & 0 & 0 & 0 & 0 & 0 & 0 & 0 & 0 & 0 & 0 & 0 & 0 & j & 0 & 0 & 0 & 0 & 0 & 0 & 0 & -1 & 0 & 0 \\
0 & 0 & 1 & 0 & 0 & 0 & 0 & 0 & 0 & 0 & 0 & 0 & 0 & 0 & 0 & j & 0 & 0 & 0 & 0 & 0 & 0 & 0 & 0 & 0 & 0 & 0 & 0 & 0 & 0 & 0 & 0 \\
0 & 0 & 0 & 0 & 0 & 0 & 0 & 0 & 0 & 0 & 0 & 0 & 0 & 0 & 0 & 0 & -1 & 0 & 0 & 0 & 0 & 0 & j & 0 & 0& 0 & 0 & 0 & 0 & 0 & 0 & 0 \\
0 & 0 & 0 & 0 & 0 & -1 & 0 & 0 & 0 & j & 0 & 0 & 0 & 0 & 0 & 0 & 0 & 0 & 0 & 0 & 0 & 0 & 0 & 0 & 0& 0 & 0 & 0 & 0 & 0 & 0 & 0 \\
0 & 0 & 0 & 0 & 0 & 0 & 0 & 0 & 0 & 0 & 0 & 0 & 0 & 0 & 0 & 0 & 0 & j & 0 & 0 & 0 & 0 & 0 & 0 & 0 & 1 & 0 & 0 & 0 & 0 & 0 & 0 \\
0 & 0 & 0 & 1 & 0 & 0 & 0 & 0 & 0 & 0 & 0 & 0 & 0 & -j & 0 & 0 & 0 & 0 & 0 & 0 & 0 & 0 & 0 & 0 & 0& 0 & 0 & 0 & 0 & 0 & 0 & 0 \\
0 & 0 & 0 & 0 & 0 & 0 & 0 & 0 & 0 & 0 & 0 & 0 & 0 & 0 & 0 & 0 & 0 & 0 & 0 & 0 & 0 & 0 & 0 & j & 0 & 0 & 0 & 0 & -1 & 0 & 0 & 0 \\
0 & 0 & 0 & 0 & 0 & 0 & 1 & 0 & 0 & 0 & j & 0 & 0 & 0 & 0 & 0 & 0 & 0 & 0 & 0 & 0 & 0 & 0 & 0 & 0 & 0 & 0 & 0 & 0 & 0 & 0 & 0 \\
0 & 0 & 0 & 0 & 0 & 0 & 0 & 0 & 0 & 0 & 0 & 0 & 0 & 0 & 0 & 0 & 0 & 0 & 0 & j & 0 & 0 & 0 & 0 & 0 & 0 & 0 & 1 & 0 & 0 & 0 & 0 \\
\end{bsmallmatrix*}
.
\end{align}

{\small
\singlespacing
\bibliographystyle{siam}
\bibliography{reference,bibcleanoutput}
}

\end{document}

%% file: 32_ULA_URA_OV_1024_2.pstex_t
\begin{picture}(0,0)%
\includegraphics{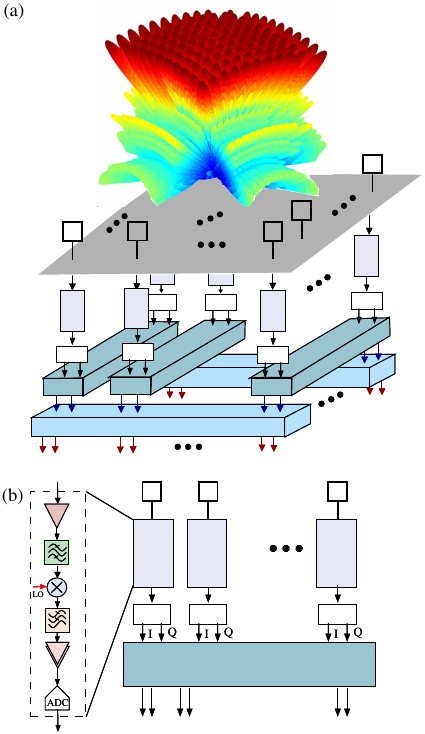}%
\end{picture}%
\setlength{\unitlength}{3947sp}%
\begingroup\makeatletter\ifx\SetFigFont\undefined%
\gdef\SetFigFont#1#2#3#4#5{%
  \reset@font\fontsize{#1}{#2pt}%
  \fontfamily{#3}\fontseries{#4}\fontshape{#5}%
  \selectfont}%
\fi\endgroup%
\begin{picture}(3398,5858)(-4514,-11144)
\put(-3254,-7718){\makebox(0,0)[lb]{\smash{{\SetFigFont{5}{6.0}{\rmdefault}{\mddefault}{\updefault}{\color[rgb]{0,0,0}HT}%
}}}}
\put(-2780,-7721){\makebox(0,0)[lb]{\smash{{\SetFigFont{5}{6.0}{\rmdefault}{\mddefault}{\updefault}{\color[rgb]{0,0,0}HT}%
}}}}
\put(-3254,-7718){\makebox(0,0)[lb]{\smash{{\SetFigFont{5}{6.0}{\rmdefault}{\mddefault}{\updefault}{\color[rgb]{0,0,0}HT}%
}}}}
\put(-2780,-7721){\makebox(0,0)[lb]{\smash{{\SetFigFont{5}{6.0}{\rmdefault}{\mddefault}{\updefault}{\color[rgb]{0,0,0}HT}%
}}}}
\put(-3336,-10217){\makebox(0,0)[lb]{\smash{{\SetFigFont{5}{6.0}{\rmdefault}{\mddefault}{\updefault}{\color[rgb]{0,0,0}HT}%
}}}}
\put(-2886,-10217){\makebox(0,0)[lb]{\smash{{\SetFigFont{5}{6.0}{\rmdefault}{\mddefault}{\updefault}{\color[rgb]{0,0,0}HT}%
}}}}
\put(-3336,-10217){\makebox(0,0)[lb]{\smash{{\SetFigFont{5}{6.0}{\rmdefault}{\mddefault}{\updefault}{\color[rgb]{0,0,0}HT}%
}}}}
\put(-2886,-10217){\makebox(0,0)[lb]{\smash{{\SetFigFont{5}{6.0}{\rmdefault}{\mddefault}{\updefault}{\color[rgb]{0,0,0}HT}%
}}}}
\put(-3967,-8135){\makebox(0,0)[lb]{\smash{{\SetFigFont{5}{6.0}{\rmdefault}{\mddefault}{\updefault}{\color[rgb]{0,0,0}HT}%
}}}}
\put(-3444,-8115){\makebox(0,0)[lb]{\smash{{\SetFigFont{5}{6.0}{\rmdefault}{\mddefault}{\updefault}{\color[rgb]{0,0,0}HT}%
}}}}
\put(-2361,-8138){\makebox(0,0)[lb]{\smash{{\SetFigFont{5}{6.0}{\rmdefault}{\mddefault}{\updefault}{\color[rgb]{0,0,0}HT}%
}}}}
\put(-1612,-7707){\makebox(0,0)[lb]{\smash{{\SetFigFont{5}{6.0}{\rmdefault}{\mddefault}{\updefault}{\color[rgb]{0,0,0}HT}%
}}}}
\put(-3967,-8135){\makebox(0,0)[lb]{\smash{{\SetFigFont{5}{6.0}{\rmdefault}{\mddefault}{\updefault}{\color[rgb]{0,0,0}HT}%
}}}}
\put(-3444,-8115){\makebox(0,0)[lb]{\smash{{\SetFigFont{5}{6.0}{\rmdefault}{\mddefault}{\updefault}{\color[rgb]{0,0,0}HT}%
}}}}
\put(-2361,-8138){\makebox(0,0)[lb]{\smash{{\SetFigFont{5}{6.0}{\rmdefault}{\mddefault}{\updefault}{\color[rgb]{0,0,0}HT}%
}}}}
\put(-1612,-7707){\makebox(0,0)[lb]{\smash{{\SetFigFont{5}{6.0}{\rmdefault}{\mddefault}{\updefault}{\color[rgb]{0,0,0}HT}%
}}}}
\put(-2849,-10636){\makebox(0,0)[lb]{\smash{{\SetFigFont{7}{8.4}{\rmdefault}{\mddefault}{\updefault}{\color[rgb]{0,0,0}N-point ADFT}%
}}}}
\put(-1850,-10217){\makebox(0,0)[lb]{\smash{{\SetFigFont{5}{6.0}{\rmdefault}{\mddefault}{\updefault}{\color[rgb]{0,0,0}HT}%
}}}}
\put(-1850,-10217){\makebox(0,0)[lb]{\smash{{\SetFigFont{5}{6.0}{\rmdefault}{\mddefault}{\updefault}{\color[rgb]{0,0,0}HT}%
}}}}
\put(-2104,-8167){\rotatebox{30.0}{\makebox(0,0)[lb]{\smash{{\SetFigFont{5}{6.0}{\rmdefault}{\mddefault}{\updefault}{\color[rgb]{0,0,0}\scalebox{0.9}{N-point ADFT}}%
}}}}}
\put(-3223,-8167){\rotatebox{30.0}{\makebox(0,0)[lb]{\smash{{\SetFigFont{5}{6.0}{\rmdefault}{\mddefault}{\updefault}{\color[rgb]{0,0,0}\scalebox{0.9}{N-point ADFT}}%
}}}}}
\put(-3500,-8727){\makebox(0,0)[lb]{\smash{{\SetFigFont{6}{7.2}{\rmdefault}{\mddefault}{\updefault}{\color[rgb]{0,0,0}N-point ADFT}%
}}}}
\end{picture}%

%% file: Fig_sim_algo1.pstex_t
\begin{picture}(0,0)%
\includegraphics{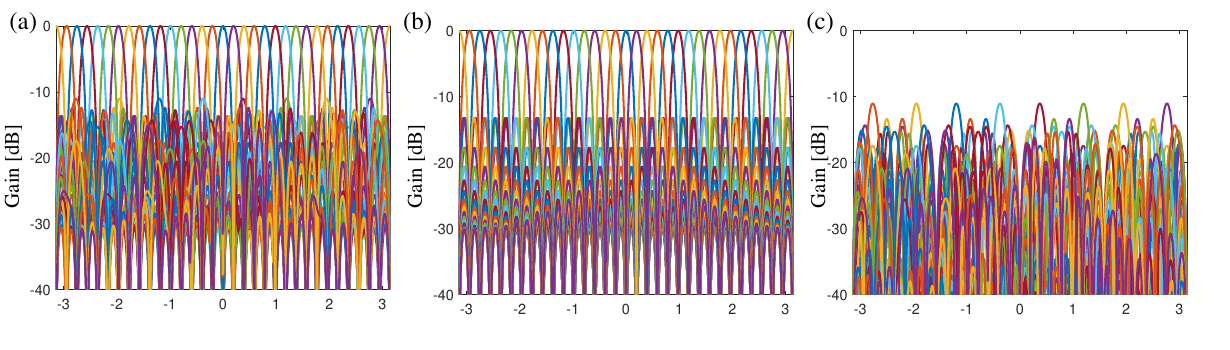}%
\end{picture}%
\setlength{\unitlength}{3947sp}%
\begingroup\makeatletter\ifx\SetFigFont\undefined%
\gdef\SetFigFont#1#2#3#4#5{%
  \reset@font\fontsize{#1}{#2pt}%
  \fontfamily{#3}\fontseries{#4}\fontshape{#5}%
  \selectfont}%
\fi\endgroup%
\begin{picture}(9824,2785)(1201,-1946)
\put(2938,-1846){\makebox(0,0)[lb]{\smash{{\SetFigFont{12}{14.4}{\rmdefault}{\mddefault}{\updefault}{\color[rgb]{0,0,0}$\omega$}%
}}}}
\put(6168,-1851){\makebox(0,0)[lb]{\smash{{\SetFigFont{12}{14.4}{\rmdefault}{\mddefault}{\updefault}{\color[rgb]{0,0,0}$\omega$}%
}}}}
\put(9321,-1873){\makebox(0,0)[lb]{\smash{{\SetFigFont{12}{14.4}{\rmdefault}{\mddefault}{\updefault}{\color[rgb]{0,0,0}$\omega$}%
}}}}
\end{picture}%

%% file: worst_bins.pstex_t
\begin{picture}(0,0)%
\includegraphics{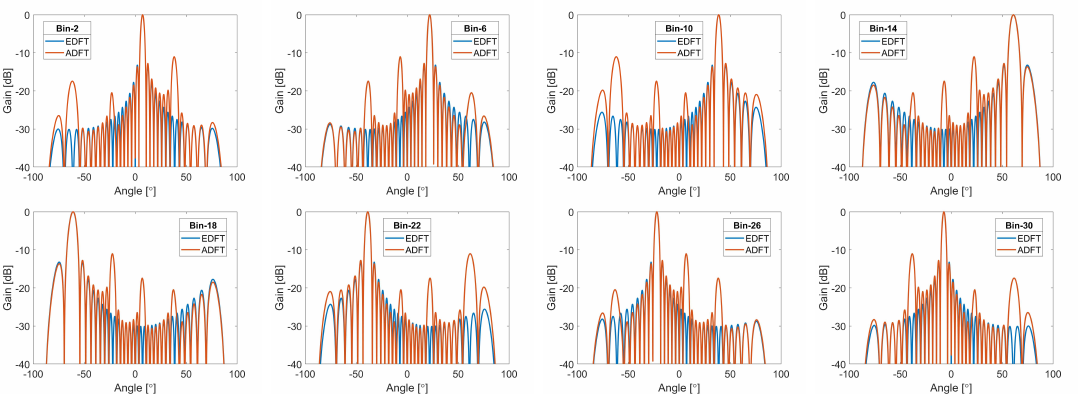}%
\end{picture}%
\setlength{\unitlength}{3947sp}%
\begingroup\makeatletter\ifx\SetFigFont\undefined%
\gdef\SetFigFont#1#2#3#4#5{%
  \reset@font\fontsize{#1}{#2pt}%
  \fontfamily{#3}\fontseries{#4}\fontshape{#5}%
  \selectfont}%
\fi\endgroup%
\begin{picture}(8621,3148)(76,-2534)
\end{picture}%

%% file: ULA_URA_sim_beams_R1.pstex_t
\begin{picture}(0,0)%
\includegraphics{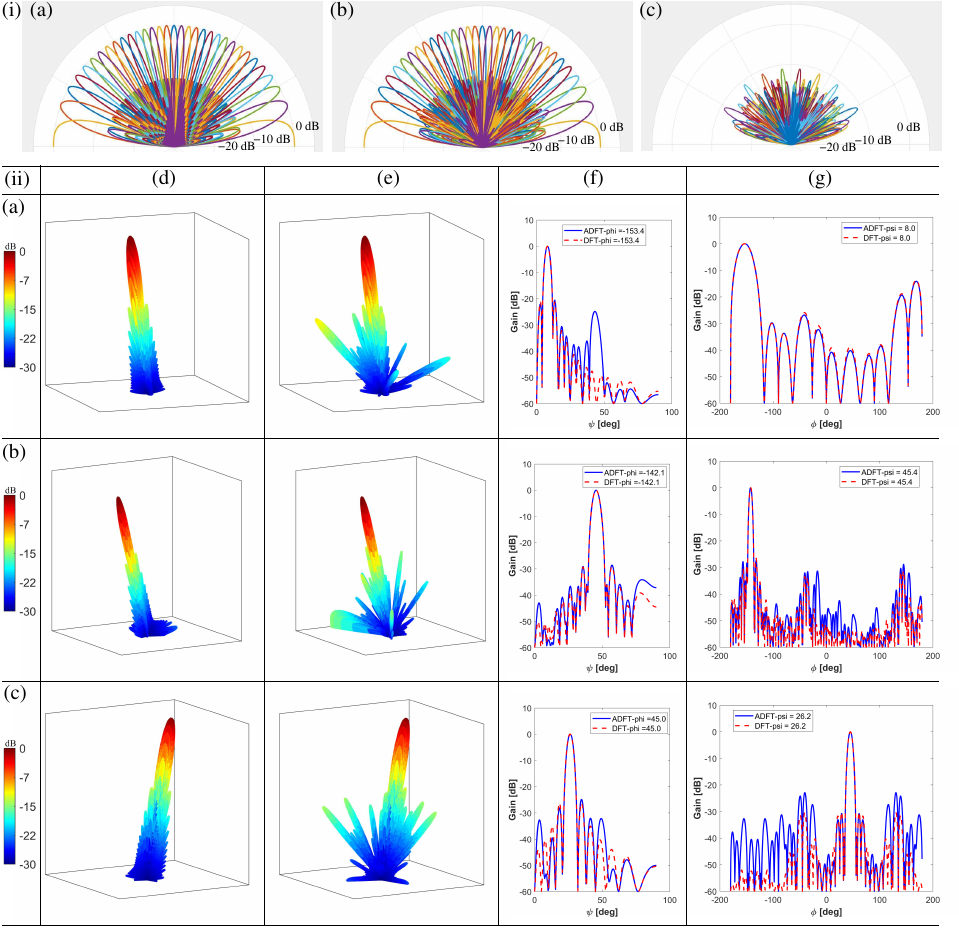}%
\end{picture}%
\setlength{\unitlength}{3947sp}%
\begingroup\makeatletter\ifx\SetFigFont\undefined%
\gdef\SetFigFont#1#2#3#4#5{%
  \reset@font\fontsize{#1}{#2pt}%
  \fontfamily{#3}\fontseries{#4}\fontshape{#5}%
  \selectfont}%
\fi\endgroup%
\begin{picture}(7665,7410)(1186,-6823)
\put(8442,-593){\makebox(0,0)[lb]{\smash{{\SetFigFont{6}{7.2}{\rmdefault}{\mddefault}{\updefault}{\color[rgb]{0,0,0}$90^{\circ}$}%
}}}}
\put(6326,-605){\makebox(0,0)[lb]{\smash{{\SetFigFont{6}{7.2}{\rmdefault}{\mddefault}{\updefault}{\color[rgb]{0,0,0}$-90^{\circ}$}%
}}}}
\put(3575,-606){\makebox(0,0)[lb]{\smash{{\SetFigFont{6}{7.2}{\rmdefault}{\mddefault}{\updefault}{\color[rgb]{0,0,0}$90^{\circ}$}%
}}}}
\put(1359,-611){\makebox(0,0)[lb]{\smash{{\SetFigFont{6}{7.2}{\rmdefault}{\mddefault}{\updefault}{\color[rgb]{0,0,0}$-90^{\circ}$}%
}}}}
\put(6010,-606){\makebox(0,0)[lb]{\smash{{\SetFigFont{6}{7.2}{\rmdefault}{\mddefault}{\updefault}{\color[rgb]{0,0,0}$90^{\circ}$}%
}}}}
\put(3834,-606){\makebox(0,0)[lb]{\smash{{\SetFigFont{6}{7.2}{\rmdefault}{\mddefault}{\updefault}{\color[rgb]{0,0,0}$-90^{\circ}$}%
}}}}
\put(7519,434){\makebox(0,0)[lb]{\smash{{\SetFigFont{6}{7.2}{\rmdefault}{\mddefault}{\updefault}{\color[rgb]{0,0,0}$0^{\circ}$}%
}}}}
\put(2569,414){\makebox(0,0)[lb]{\smash{{\SetFigFont{6}{7.2}{\rmdefault}{\mddefault}{\updefault}{\color[rgb]{0,0,0}$0^{\circ}$}%
}}}}
\put(5014,425){\makebox(0,0)[lb]{\smash{{\SetFigFont{6}{7.2}{\rmdefault}{\mddefault}{\updefault}{\color[rgb]{0,0,0}$0^{\circ}$}%
}}}}
\end{picture}%

%% file: overview_rf.pstex_t
\begin{picture}(0,0)%
\includegraphics{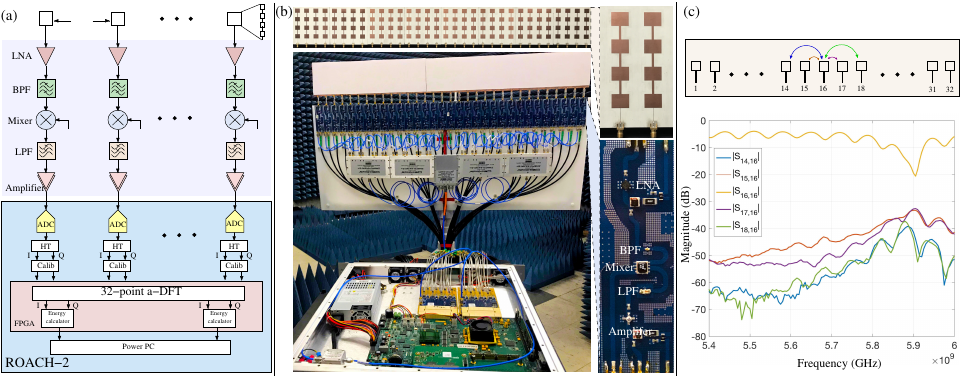}%
\end{picture}%
\setlength{\unitlength}{1895sp}%
\begingroup\makeatletter\ifx\SetFigFont\undefined%
\gdef\SetFigFont#1#2#3#4#5{%
  \reset@font\fontsize{#1}{#2pt}%
  \fontfamily{#3}\fontseries{#4}\fontshape{#5}%
  \selectfont}%
\fi\endgroup%
\begin{picture}(16002,6259)(661,-20998)
\put(1865,-15101){\makebox(0,0)[lb]{\smash{{\SetFigFont{5}{6.0}{\rmdefault}{\mddefault}{\updefault}{\color[rgb]{0,0,0}$\Delta x$}%
}}}}
\put(1726,-17011){\makebox(0,0)[lb]{\smash{{\SetFigFont{5}{6.0}{\rmdefault}{\mddefault}{\updefault}{\color[rgb]{0,0,0}$f_{lo}$}%
}}}}
\put(2926,-17011){\makebox(0,0)[lb]{\smash{{\SetFigFont{5}{6.0}{\rmdefault}{\mddefault}{\updefault}{\color[rgb]{0,0,0}$f_{lo}$}%
}}}}
\put(4876,-17011){\makebox(0,0)[lb]{\smash{{\SetFigFont{5}{6.0}{\rmdefault}{\mddefault}{\updefault}{\color[rgb]{0,0,0}$f_{lo}$}%
}}}}
\end{picture}%

%% file: mes_beams.pstex_t
\begin{picture}(0,0)%
\includegraphics{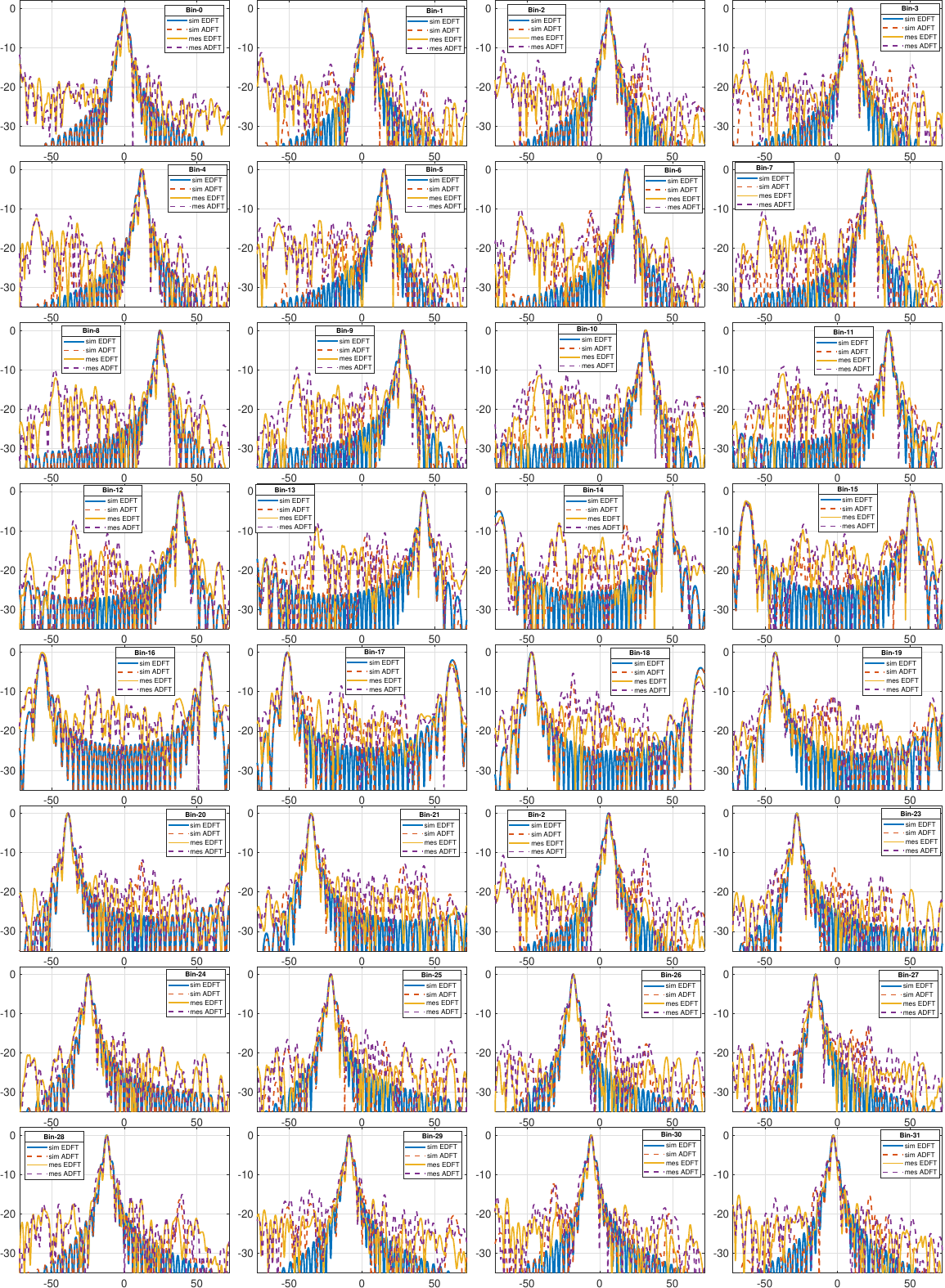}%
\end{picture}%
\setlength{\unitlength}{3947sp}%
\begingroup\makeatletter\ifx\SetFigFont\undefined%
\gdef\SetFigFont#1#2#3#4#5{%
  \reset@font\fontsize{#1}{#2pt}%
  \fontfamily{#3}\fontseries{#4}\fontshape{#5}%
  \selectfont}%
\fi\endgroup%
\begin{picture}(9225,12598)(76,-11984)
\end{picture}%

%% file: Fig_separation.pstex_t
\begin{picture}(0,0)%
\includegraphics{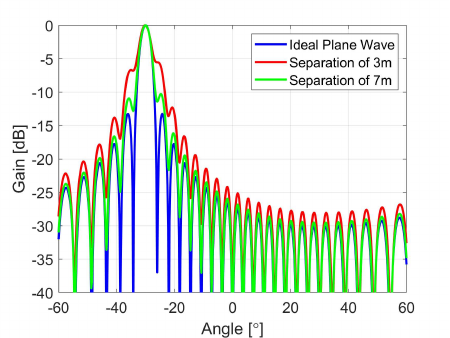}%
\end{picture}%
\setlength{\unitlength}{4144sp}%
\begingroup\makeatletter\ifx\SetFigFont\undefined%
\gdef\SetFigFont#1#2#3#4#5{%
  \reset@font\fontsize{#1}{#2pt}%
  \fontfamily{#3}\fontseries{#4}\fontshape{#5}%
  \selectfont}%
\fi\endgroup%
\begin{picture}(3418,2565)(2251,-4471)
\end{picture}%

%% file: mes_sim_2D_beams.pstex_t
\begin{picture}(0,0)%
\includegraphics{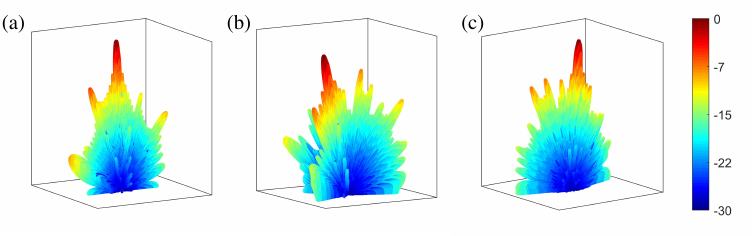}%
\end{picture}%
\setlength{\unitlength}{3947sp}%
\begingroup\makeatletter\ifx\SetFigFont\undefined%
\gdef\SetFigFont#1#2#3#4#5{%
  \reset@font\fontsize{#1}{#2pt}%
  \fontfamily{#3}\fontseries{#4}\fontshape{#5}%
  \selectfont}%
\fi\endgroup%
\begin{picture}(6015,1868)(1261,-6729)
\end{picture}%